\def\preprint{1}			

\ifdefined\preprint
\documentclass[preprint,review,12pt]{elsarticle}
\fi
\ifdefined\wordcount
\documentclass[final,3p,times,twocolumn]{elsarticle}
\fi
\ifdefined\final
\documentclass[final,3p,times,twocolumn]{elsarticle}
\fi

\DeclareMathSizes{7}{7}{6}{6}

\usepackage{graphicx,stfloats}
\usepackage{color}

\usepackage[fleqn]{amsmath} 
\usepackage{hyperref} 
\usepackage{soul} 

\usepackage{tabularx}
\newcolumntype{Y}{>{\centering\arraybackslash}X}
\graphicspath{{./figs/}}
\usepackage{cleveref}
\usepackage{tikz}
\usepackage{natbib}
\usepackage[separate-uncertainty = true]{siunitx}
\usepackage{nicefrac}
\usepackage{caption}
\usepackage{subcaption}
\usepackage{adjustbox}
 \usepackage{multirow}
\usepackage{wasysym}
\usepackage{gensymb}
\usepackage{amsthm}

\newcommand\stoi{\mathrm{st}}
\newcommand\soot{\mathrm{soot}}

\biboptions{sort&compress}

\makeatletter
\def\ps@pprintTitle{%
  \let\@oddhead\@empty
  \let\@evenhead\@empty
  \let\@oddfoot\@empty
  \let\@evenfoot\@oddfoot
}
\makeatother

\begin{document}

\begin{frontmatter}

\title{Large Eddy Simulation of the evolution of the soot size distribution in turbulent nonpremixed flames using the Bivariate Multi-Moment Sectional Method}

\author[fir]{Hernando Maldonado Colmán\corref{cor1}}
\ead{hm2524@princeton.edu}

\author[fir]{Michael E. Mueller}

\address[fir]{Department of Mechanical and Aerospace Engineering, Princeton University, Princeton, NJ 08544, USA}
\cortext[cor1]{Corresponding author.}

\begin{abstract}
	
\par
A joint volume-surface formalism of the Multi-Moment Sectional Method (MMSM) is developed to describe the evolution of soot size distribution in turbulent reacting flows. The bivariate MMSM (or BMMSM) considers three statistical moments per section, including the total soot number density, total soot volume, and total soot surface area per section. A linear profile along the volume coordinate is considered to reconstruct the size distribution within each section, which weights a delta function along the surface coordinate. The closure for the surface considers that the primary particle diameter is constant so the surface/volume ratio constant within each section. The inclusion of the new variable in BMMSM allows for the description of soot's fractal aggregate morphology compared to the strictly spherical assumption of its univariate predecessor. BMMSM is shown to reproduce bimodal soot size distributions in simulations of one-dimensional laminar sooting flames as in experimental measurements.  To demonstrate its performance in turbulent reacting flows, BMMSM is coupled to a Large Eddy Simulation (LES) framework to simulate a laboratory-scale turbulent nonpremixed jet flame. Computational results are validated against available experimental measurements of soot size distribution, showing the ability of BMMSM to reproduce the evolution of the size distribution from unimodal to bimodal moving downstream in the flame. In general, varying the number of sections has limited influence on results, and accurate results are obtained with as few as eight sections so 24 total degrees of freedom. The impact of using a different statistical model for soot, the Hybrid Method of Moment (HMOM), is also investigated. Aside from the fact that HMOM cannot provide information about the soot size distribution, the most significant qualitative difference between HMOM and BMMSM is in number density in the oxidation region of the flame, suggesting that BMMSM outperforms HMOM in reproducing key aspects of the soot oxidation process. Finally, the total computational cost of using BMMSM as low as approximately 44\% more than the cost of HMOM. Therefore, the new formulation results in a computationally efficient approach for the soot size distribution in turbulent reacting flows, enabling simulations of the soot size distribution in complex industrial configurations that are unattainable using traditional sectional models.

\end{abstract}


\end{frontmatter}

\textbf{Novelty and Significance Statement}

This paper introduces a joint volume-surface formulation of the Multi-Moment Sectional Method (MMSM), called the bivariate MMSM (BMMSM), which allows for tracking the soot size distribution in turbulent reacting flows including soot's fractal aggregate morphology. Also, this model has been implemented in a Large Eddy Simulation framework, which allows for the description of the evolution of the soot size distribution in turbulent combustion. BMMSM is shown to qualitatively and quantitatively predict the evolution of the soot size distribution in laminar and turbulent flames. Overall, the total computational cost for turbulent reacting flows is only marginally more (44\%) than the cost of traditional moment methods, which do not provide the soot size distribution.

\textbf{Authors contribution}

\underline{HMC}: performed research, software development, writing -- original draft \& editing

\underline{MEM}: designed research, writing -- review \& editing, project administration


\ifdefined \wordcount
\clearpage
\fi

\section{Introduction}
\label{sec:introduction}

The need for clean, efficient combustion devices is imperative to prevent emissions from propulsion and power systems. In this context, mitigating pollutants is crucial due to their negative effects on the environment and living organisms, which includes addressing the formation of soot particles. Soot formation is a significant concern for engineers and researchers, for fine soot particles directly affect air quality (hazardous for the respiratory system) and can contribute to global warming \cite{LMM2023SpecialIssue}. However, tracking the evolution of a tremendous number of soot particles in turbulent reacting flows is a grand experimental and computational challenge.  In computational modeling, this would entail of a mathematical framework capable of handling a transport equation for individual soot particles. For that reason, statistical soot models are required in simulations, which amounts to tracking the evolution of the soot Number Density Function (NDF). Its evolution is governed by a Population Balance Equation (PBE). For instance, Monte Carlo solves for the evolution of a downsampled representative set of particles \cite{balthasar2003stochastic}. However, Monte Carlo is prohibitively expensive beyond 0D or 1D systems so would not be suitable for large-scale industrial systems. A computationally efficient approach for solving the PBE is through the transport of a few statistical moments of the NDF, known as the method of moments~\cite{frenklach1987aerosol}. These statistical models can accurately represent global soot attributes from the NDF, such as soot volume fraction or number density, but they are unable to track the detailed evolution of the size distribution.

Computationally, the traditional approach for modeling the evolution of the soot size distribution is the sectional method. In the sectional method, the size distribution is subdivided into a set of ``sections'' that are governed by statistical physics of particle dynamics and surface reactivity with the surrounding gas \cite{park2004novel}. In traditional soot sectional methods, the NDF evolution is described by transporting the number density per soot section, which is often defined with respect to discrete volume intervals. The primary computational challenge with the sectional method is cost since it typically requires a range of 30 to 100 sections to accurately track the soot size distribution. Additionally, these volume-only models can only consider spherical soot and cannot characterize fractal aggregate morphology. While some approaches attempt to model a fixed size at which aggregation begins to occur \cite{kazakov1998dynamic,bhatt2009analysis,netzell2007calculating,rodrigues2018coupling,schiener2019transported}, the use of multivariate sections directly addresses this issue of morphology without ad hoc size limits. For instance, Zhang et al.~\cite{zhang2009modeling,zhang2009numerical} developed a sectional method considering two internal coordinates within soot mass-based sections, namely number densities of soot primary particles and soot aggregates. However, given the added model fidelity and improved accuracy, such an investment in higher computational resources is not without merit. Rather than adding unknowns per section for morphology considerations, other approaches instead add unknowns per section to account for different chemical reactivity (${\rm H}/{\rm C}$ ratio) within each section \cite{saggese2015kinetic}.

While the multivariate sectional methods aforementioned are simply too expensive for simulating turbulent sooting flames, simplified sectional models have been successful in reproducing the evolution of the soot size distribution in simulations of turbulent sooting flames \cite{sewerin2018pbe, rodrigues2018coupling, schiener2019transported, huo2022coupled}, which may consider directly spherical soot or including the aggregation limit occurring at a critical size. 


Other (non-sectional) methodologies have also been extended to include the soot size distribution via mathematical reconstruction of the NDF. For instance, the Extended Quadrature Method of Moments for soot \cite{salenbauch2015modeling,wick2017modeling} was implemented in an LES framework \cite{cokuslu2022soot,ferraro2022soot} and allowed the reconstruction of the NDF by approximating them using weighted kernel distribution functions (KDFs). Despite this, the model was not validated against experimental data of turbulent flames \cite{ferraro2022soot}. A three-equation model for soot formation was introduced by Franzelli et al.~\cite{franzelli2019three}, in which global soot quantities are transported. Similar to EQMOM, the NDF was retrieved by approximating it as weighted Pareto distributions, which was numerically validated against a sectional method in laminar and turbulent flames. The open question with these mathematical reconstruction is accuracy of the resulting size distribution since it does not rely on the underlying PBE.



To address the computational cost shortcomings of traditional sectional methods, Yang and Mueller \cite{yang2019multi} introduced a Multi-Moment Sectional Method (MMSM) that essentially combines a sectional method with the method of moments. Within each section, the approach reconstructs the NDF using a linear distribution whose parameters are defined using two section-local moments. MMSM has a lower computational cost than a traditional sectional method since fewer sections and overall degrees of freedom are required, due to an improved rate of convergence compared to traditional sectional methods \cite{yang2019multi}. This implies that MMSM can be as accurate but faster than traditional sectional methods. However, fractal aggregate morphology was not fully described since previous work only considered a univariate (strictly spherical) formulation, failing to capture bimodal soot size distributions in laminar flames. Additionally, MMSM was not applied to turbulent flames

In this work, a bivariate formulation of MMSM is developed, called BMMSM, which introduces a joint volume-surface description of soot. Following the previous work of the senior author's group, BMMSM includes a bivariate formulation based on the formulation of the Hybrid Method of Moments (HMOM) \cite{mueller2009hybrid}. BMMSM is first tested in 1D laminar sooting flames and compared to experimental measurements. BMMSM is then integrated in an LES framework in order to simulate a laboratory-scale turbulent nonpremixed jet flame. Combustion and subfilter closures are adapted from Ref.~\cite{hmcolman2023from}, which were initially developed for HMOM. Computational results are analyzed and compared against available experimental data of soot size distributions. The LES results with BMMSM are then analyzed to provide better understanding of key aspects of the evolution of the soot size distribution. Finally, the computational cost of turbulent combustion simulations with BMMSM is then compared to simulations using HMOM.

\section{The Bivariate Multi-Moment Sectional Method}
\label{sec:bmmsm}

The univariate Multi-Moment Sectional Method developed by Yang and Mueller \cite{yang2019multi} considered strictly spherical soot particles described by the soot particle volume $V$. A statistical representation of the particle size distribution was considered through a NDF $n(V)$, which was discretized in volume sections $V_{i}$. In each section, two local (up to first-order) moments were solved, the number density $M_{0}^{i}$ and the volume moments $M_{1}^{i}$, defined as
\begin{equation}
    M_{x}^{i}=\int_{\Omega_{i}^{V}}n_{i}(V)V^{x}{\rm d}V,
\end{equation}
\noindent
where $\Omega_{i}^{V}$ indicates the support of the $i^{th}$ section (out of $N_{s}$). The local moments were used to reconstruct the local size distribution with linear approximation within each section. The section size $\Delta V_{i}$, except for the last section, was centered at $V_{i}$. The total number of variables $N_{v}^{\rm s}$ considered in this approach was $N_{v}=2N_{s}$. In the following subsections, the MMSM formalism is extended to a bivariate soot model.

\subsection{Soot governing equations}

The bivariate MMSM, hereafter referred as BMMSM, accounts for a joint volume-surface area formalism and can describe the fractal aggregate morphology of soot particles \cite{mueller2009joint}. The NDF $n(V,S)$ is then a function of two internal coordinates: the soot particle volume $V$ and surface area $S$. Mueller et al.~\cite{mueller2009hybrid} have shown computationally, through Monte Carlo simulations, that the joint surface-volume NDF is divided in two regions with fundamentally different $S-V$ distributions: spherical particles ($S \propto V^{2/3}$) and fractal aggregates. Therefore, while a global relationship between $S$ and $V$ is not expected to be accurate or general, section-local relationships between $S$ and $V$, allowed to vary based on the dynamics of the PBE, are expected to be accurate, and a fully multidimensional MMSM is not required. Given this assumption, the bivariate NDF $n(V,S)$ in section $i$ is modeled with a presumed dependence of $S$ and $V$ within the section:
\begin{equation}
    n(V,S)=n(V)\delta(S-\hat{S}_{i}(V)),
\end{equation}
\noindent
where $\hat{S}(V)$ is the soot surface area for a soot particle of volume $V$ in the $i^{th}$ section, which needs closure. Then, the bivariate moment in the $i^{th}$ section is defined as
\begin{equation}
    \begin{aligned}
        M_{x,y}^{i}&=\iint_{\Omega_{i}^{V,S}}n(V,S)V^{x}S^{y}{\rm d}V{\rm d}S \\
                   &=\int_{\Omega_{i}^{V}}n(V)V^{x}\hat{S}_{i}^{y}(V){\rm d}V.
    \label{eq:mxy_def}
    \end{aligned}
\end{equation}
\noindent
As a first approximation, the primary particle diameter $d_{p}=6V/S$ is assumed constant within each section, that is, $S/V$ remains constant within each section and equal to $\alpha_{i}$. The model variable $\alpha_{i}$ estimates the proportionality $S/V$ and is an index of the sphericity or degree of aggregation of the soot morphology in a given section. Low $\alpha_{i}$ values indicate that soot approaches a spherical shape, with lower limit $(36\pi/V_{i})^{1/3}$. To ensure nucleated particles are spherical \cite{frenklach1991detailed}, the first section considers $\alpha_{1}=(36\pi/V_{1})^{1/3}$ and is centered at $V_{1}=V_{0}$, which corresponds to  the nucleated soot particle volume $V_{0}$. Then, the soot surface area $\hat{S}_{i}(V)$ of a soot particle of volume $V$ in the $i^{th}$ section can be approximated as
\begin{equation}
    \hat{S}_{i}(V)=\frac{M_{0,1}^{i}}{M_{1,0}^{i}}V=\alpha_{i}V,
    \label{eq:hatSV}
\end{equation}
\noindent
which, inserted to  Eq.~\ref{eq:mxy_def}, leads to
\begin{equation}
    M_{x,y}^{i}=\alpha_{i}^{y}\int_{\Omega_{i}^{V}}n(V)V^{x+y}{\rm d}V=\alpha_{i}^{y}M_{x+y,0}^{i}.
\end{equation}
This new closure requires an additional degree of freedom per section: the total surface moment $M_{0,1}^{i}$. This increases the total number of unknowns $N_{v}^{\rm s}$ to $N_{v}=3N_{s}$.

The spatiotemporal and size ($V$-$S$) evolution of the NDF is governed by the Population Balance Equation. In the context of the Multi-Moment Sectional Method, the PBE is integrated to obtain the moment transport equations of sections $i=1,\dots,N_{s}$, which leads to
\begin{equation}
    \frac{\partial M_{x,y}^{i}}{\partial t} + \frac{\partial u_{j}^{*}M_{x,y}^{i}}{\partial x_{j}}=\dot{M}_{x,y}^{i},
    \label{eq:mom_gov_eqs}
\end{equation}
\noindent
where $u_{j}^{*}$ is the total velocity, including flow and thermophoretic effects \cite{waldmann1966thermophoresis}, of diffusionless soot particles \cite{bisetti2012formation}, and $\dot{M}_{x,y}^{i}$ are the soot source terms, whose closures are developed in subsection \ref{subsec:soot_source_terms}.

\subsection{Local volume-dependent NDF reconstruction}
The bivariate NDF defined above needs closure for the volume-dependent NDF $n(V)$ within each section. The closure chosen by Yang and Mueller \cite{yang2019multi} is retained. Here, $n(V)$ is modeled with a linear distribution in the first $N_{s}-1$ sections as follows:
\begin{equation}
    \Delta V_{i}n(V)|_{\Omega_{i}}=a(V-V_{i})+b,
    \label{eq:nV_approx}
\end{equation}
\noindent
where $V_{i}$ is the $i^{th}$ section with section size $\Delta V_{i}$ along the volume coordinate and $a$ and $b$ are local moment-dependent model parameters, defined as
\begin{equation}
    a=\frac{12}{(\Delta V_{i})^{2}}\left(M_{1,0}^{i}-M_{0,0}^{i}V_{i}\right)
\end{equation}
and
\begin{equation}
    b=M_{0,0}^{i},
    \label{eq:b_par}
\end{equation}
\noindent
respectively. In the last section ($i=N_{s})$, an exponential distribution is considered:
\begin{equation}
    \left.\Delta V_{N_{s}} n(V)\right|_{\Omega_{N_{s}}} =
    \Delta V_{N_{s}} b a \exp \left\{-a\left[V-\left(V_{N_{s}}-\frac{\Delta V_{N_{s}}}{2}\right)\right]\right\},
\end{equation}
\noindent
with $a$ as
\begin{equation}
    a=\frac{M_{0,0}^{N_{s}}}{M_{1,0}^{N_{s}} - 2V_{N_{s}} M_{0,0}^{N_{s}}/(f_{s}+1)}
\end{equation}
\noindent
and $b$ the same as in Eq.~\ref{eq:b_par}. 
The model considers a geometrically increasing section size $\Delta V_{i}$, to reproduce rapid coagulation processes, with a spacing factor $f_{s}$ defined as 
\begin{equation}
f_{s}=\frac{\Delta V_{i+1}}{\Delta V_{i}},
\end{equation}
leading to
\begin{equation}
\Delta V_{i}=2V_{i}\left(\frac{f_{s}-1}{f_{s}+1}\right).
\end{equation}
The section support in volume space is $\Omega_{i}=[V_{i}-\Delta V_{i}/2,V_{i}+\Delta V_{i}/2)$ in the first $N_{s}-1$ sections (finite), while in the last section is $\Omega_{N_{s}}=[V_{N_{s}}-\Delta V_{N_{s}}/2,\infty)$ (semi-infinite). In the last section, the true barycenter is
\begin{equation}
V_{N_{s}}^{*}=\frac{2V_{N_{s}}}{f_{s}+1}+\frac{1}{a}.    
\end{equation}

The moment source terms of Eq.~\ref{eq:mom_gov_eqs}, defined in the following subsection, are computed using the bivariate NDF and integrated as in Eq.~\ref{eq:mxy_def}. To limit the computational costs of those operations, integration using two-point Gauss-Legendre quadrature for $i<N_{s}$ and, similarly, two-point Gauss-Laguerre quadrature for $i=N_{s}$ is performed \cite{yang2019multi}.

\subsection{Source terms}
\label{subsec:soot_source_terms}

The sole term on the right-hand-side of Eq.~\ref{eq:mom_gov_eqs} represents the physiochemical phenomena affecting soot evolution: particle dynamics and soot surface reactivity with the surrounding gas. 
This work utilizes the formulations of Mueller and co-workers \cite{mueller2009hybrid,mueller2009joint,yang2019multi} with contributions from nucleation, coagulation, condensation, surface growth, and oxidation. Oxidation-induced fragmentation \cite{mueller2011modeling} is not considered in this work and is negligible in the validation and application configurations presented in subsequent sections. These source terms, previously formulated within the univariate MMSM model \cite{yang2019multi}, are extended to the bivariate formalism below.

\subsubsection{Nucleation}
In this model, a collision of a pair of PAH dimers leads to a spherical soot particle \cite{blanquart2009joint} of fixed volume $V_{0}$, which also corresponds to the first section size $V_{1}=V_{0}$. The nucleation source term is non-zero in the first section and zero otherwise and is modeled as follows:
\begin{equation}
    \dot{M}_{x,y}^{1}\rvert_{\rm nuc}=\frac{1}{2} \beta_{\rm nuc} [{\rm DIMER}]^{2} V_{0}^{x} \hat{S}^{y}(V_{0}),
\end{equation}
\noindent
where $\beta_{\rm nuc}$ is the dimer collision rate and $[{\rm DIMER}]$ is the PAH dimer concentration.

\subsubsection{Coagulation}
After a reasonably straightforward integration along the soot surface area coordinate, in similar fashion as Eq.~\ref{eq:mxy_def}, the coagulation source term is computed using a two-point Gauss-Legendre or Gauss-Laguerre quadrature in each section along the volume coordinate as in Ref.~\cite{yang2019multi}, resulting in
\begin{equation}
    \begin{aligned}
    \dot{M}_{x,y}^{i}\rvert_{\rm coag} = &\sum_{V_{(j+k)\pm} \in \Omega_{i}}^{k \leq j \leq i} \left(1-\frac{\delta_{j, k}}{2}\right) \beta_{j \pm, k \pm} V_{(j+k)\pm}^{x}S_{(j+k)\pm}^{y} w_{j \pm} N_{j \pm} w_{k \pm} N_{k \pm} \\ 
    &-w_{i \pm} N_{i \pm} \sum_{k=1}^{N_{s}} \beta_{i \pm, k \pm}V_{k \pm}^{x}S_{k \pm}^{y} w_{k \pm} N_{k \pm},
    \end{aligned}
    \label{eq:dotM_coag_eq}
\end{equation}
\noindent
where the $w_{i,\pm}$ and $V_{i,\pm}$ represent the quadrature weights and nodes, with $\pm$ as a root pair combination. Here, the  surface area $S_{i\pm}$ and number density $N_{i}\pm$ are evaluated from Eqs.~\ref{eq:hatSV} and \ref{eq:nV_approx} at $V_{i\pm}$, respectively. Then, each pair of products (evaluated with opposite root signs) is summed. Within each section, $n(V)$ is equal to the derivative of the local number density $N$ as $dN/dV=n(V)$ (see Eq.~\ref{eq:nV_approx}). $\delta_{i,j}$ is the Kronecker delta function. $\beta_{i,j}$ is the soot particle collision frequency, obtained by the harmonic mean of the collision frequencies in the free-molecular and continuum regimes \cite{pratsinis1988simultaneous}.

The collision frequency in the free-molecular regime is calculated from kinetic theory:
\begin{equation}
    \beta_{j,k}^{\rm fm} = K \left(\frac{1}{V_{j}} + \frac{1}{V_{k}}\right)^{1/2}(d_{c,j} + d_{c,k})^{2},
    \label{eq:beta_fm}
\end{equation}
where  $d_{c}$ are the collision diameters and $K$ is a constant that considers the van der Waals enhancement factor equal to 2.2, the Boltzmann constant $k_{B}$, and (constant) soot density ($\rho_{\soot}=1800\rm\:kg\:m^{-3}$) \cite{harris1988coagulation}:
\begin{equation}
    K=2.2\left( \frac{\pi k_{B} T}{2 \rho_{\soot}}\right)^{1/2}.
    \label{eq:K_cst}
\end{equation}
\noindent
The collision diameters $d_{c}$ for fractal aggregates are calculated as Kruis et al.~\cite{kruis1993simple} following the expression
\begin{equation}
    d_{c,i}=d_{p,i}n_{p,i}^{1/D_{f}}=\frac{6}{(36\pi)^{1/D_{f}}}\frac{V_{i}^{1-2/D_{f}}}{S_{i}^{1-3/D_{f}}},
    \label{eq:dc}
\end{equation}
where $d_{p,i}=6V_{i}/S_{i}$ and $n_{p,i}=V_{i}^{-2}S_{i}^{3}/(36\pi)$ are the primary particle diameter and number density in the $i^{th}$ section, respectively, and $D_{f}=1.8$ is the fractal dimension of soot particles. The collision diameter is then function of $V_{i}$ and $S_{i}=\hat{S}(V_{i})$, which is closed using Eq.~\ref{eq:hatSV}.

The collision frequency in the continuum regime is given by the Stokes-Einstein equation:
\begin{equation}
    \beta_{j,k}^{\rm cont} = \frac{2k_{B}T}{3\mu}\left(\frac{C_{j}}{d_{m,j}} + \frac{C_{k}}{d_{m,k}}\right)(d_{c,j} + d_{c,k}),
    \label{eq:beta_c}
\end{equation}
where $\mu$ is the dynamic viscosity, $d_{m,i}$ is the mobility diameter of the $i^{th}$ section, which is assumed equal to the collision diameter $d_{c,i}$ (Eq.~\ref{eq:dc}), and $C_{i}=1+1.257{\rm Kn}_{i}$ is the Cunningham slip correction factor with Knudsen number ${\rm Kn}$ based on the collision diameter $d_{c,i}$.

Finally, different collisions types are considered for $\dot{M}_{0,1}\rvert_{\rm coag}$ and are modeled as in Mueller et al.~\cite{mueller2009hybrid}, with both pure aggregation and coalescence as limits. The particle volume resulting from the collision $V_{(j+k)\pm}$ (in the production part of Eq.~\ref{eq:dotM_coag_eq}) is equal to the sum of each particle volume as it is conserved: $V_{(j+k)\pm}=V_{j\pm}+V_{k\pm}$. Regarding the surface sizes $S_{(j+k)\pm}$, if the two particles are from the first section ($j=k=1$), the pure coalescence limit is considered:
\begin{equation}
    S_{(1+1)\pm}=(36\pi)^{2/3}(V_{1}+V_{1})^{2/3}=(72\pi V_{0})^{2/3}.
    \label{eq:pure_coalescence_S}
\end{equation}
\noindent
Then, if one of the particles is from the first section and the other one from a larger section ($k=1$ and $j=i$ or $k=i$ and $j=1$, for $i>1$), then the resulting larger particle becomes slightly more spherical:
\begin{equation}
    S_{(i+1)\pm} = S_{i}+\delta S_{1}.
    \label{eq:intermediate_S}
\end{equation}
For all other collisions ($j>1$ and $k>1$), pure aggregation occurs:
\begin{equation}
    S_{(j+k)\pm} = S_{j}+S_{k}.
    \label{eq:pure_aggregation_S}
\end{equation}
In Eq.~\ref{eq:intermediate_S}, the change of surface area $\delta S_{1}$, due to the collision with a spherical particle from section 1 with volume $\delta V_{1}=V_{1}=V_{0}$, is given by the power law derived in the work of Mueller et al.~\cite{mueller2009hybrid}:
\begin{equation}
    \frac{\delta S_{1}}{S_{i}} = \frac{\delta V_{1}} {V_{i}} \left( \frac{2}{3}n_{p,i}^{-0.2043}\right),
    \label{eq:delta_S_def}
\end{equation}
where $n_{p,i}$ is the primary particle number at the $i^{th}$ section.

\subsubsection{Condensation, surface growth, and oxidation}
Source terms for soot growth, such as condensation (particle and PAH dimer collision) and surface growth (carbon addition to soot surface via the HACA mechanism \cite{frenklach1991detailed}), and soot destruction, such as oxidation (by oxygen molecule \cite{neoh1985effect} and hydroxyl radical \cite{kazakov1995detailed}), are modeled using two-point quadrature in three consecutive sections, based on a traditional sectional method of Park and Rogak \cite{park2004novel} and extended for the univariate MMSM by Yang and Mueller~\cite{yang2019multi}. The expression for $M_{0,0}^{i}$ source terms is
\begin{equation}
\begin{aligned}
    \dot{M}_{0,0}^{i}\left(I_{i}\right) &= \frac{ A_{i-1}\left[w_{(i-1) \pm} I_{(i-1) \pm} N_{(i-1) \pm}\right] }{ V_{i-1}^{*} }
    + \frac{ B_{i} \left[w_{i \pm} I_{i \pm} N_{i \pm}\right] } { V_{i}^{*} } \\
&+ \frac{ C_{i+1} \left[w_{(i+1) \pm} I_{(i+1) \pm} N_{(i+1) \pm}\right] } { V_{i+1}^{*} },
\end{aligned}
\label{eq:Mdot_00_cd_sg_ox}
\end{equation}
where the root pair of a quadrature is symbolized by $\pm$, $V_{i}^{*}=V_{i}$ for $i<N_{s}$, and, due to the barycenter variation, $V_{i}^{*}=V_{N_{s}}^{*}$ for $i=N_{s}$. $I_{i}$ represents the condensation, surface growth, and oxidation rates. The model coefficients $A_{i}$, $B_{i}$, and $C_{i}$ obey
\begin{equation}
    A_{i}+B_{i}+C_{i}=0,
    \label{eq:A_p_B_p_C_eq_0}
\end{equation}
such that the change of $M_{0,0}^{i}$ is conserved \cite{park2004novel}. The expression for $A_{i}$ is
\begin{equation}
    A_{i}=\frac{f_{s}-B_{i}(f_{s}-1)}{f_{s}^{2}-1},
    \label{eq:A_i}
\end{equation}
\noindent
and for $B_{i}$:
\begin{equation}
  B_{i}=\begin{cases}
    -\frac{1}{f_{s}+1}{\rm erf}\left( \frac{1}{4}\frac{{\rm d}\ln N_{i}}{{\rm d}\ln V_{i}}\right)& \text{if}~\frac{{\rm d}\ln N_{i}}{{\rm d}\ln V_{i}}>0,\\
    -\frac{f_{s}}{f_{s}+1}{\rm erf}\left( \frac{1}{4}\frac{{\rm d}\ln N_{i}}{{\rm d}\ln V_{i}}\right)& \text{otherwise},
  \end{cases}
  \label{eq:B_i}
\end{equation}
\noindent
which is selected to reduce numerical instability and numerical diffusion issues \cite{park2004novel}. The derivative ${\rm d}\ln N_{i}/{\rm d}\ln V_{i}$ is computed using a second-order central difference. Finally, $C_{i}$ is obtained from Eq.~\ref{eq:A_p_B_p_C_eq_0}:
\begin{equation}
    C_{i}=-(A_{i}+B_{i}).
    \label{eq:C_i}
\end{equation}
Equations \ref{eq:A_i}-\ref{eq:C_i} are valid for $i=2, \dots, N_{s}-1$. The expressions for $i=1$ and $i=N_{s}$ are derived with the intention of conserving both soot number and mass \cite{park2004novel}. $C_{1}$ is set to zero since is not needed in Eq.~\ref{eq:Mdot_00_cd_sg_ox}. Then, following Eq.~\ref{eq:C_i}, the coefficients in the first section are given by
\begin{equation}
    A_{1}=-B_{1}=\frac{1}{f_{s}-1},~C_{1}=0.
    \label{eq:A_B_C_1}
\end{equation}
\noindent
The same reasoning is made for $A_{N_{s}}$, and the model coefficients in the last section are given by
\begin{equation}
    A_{N_{s}}=0,~B_{N_{s}}=-C_{N_{s}}=\frac{f_{s}^{*}}{f_{s}^{*}-1}.
    \label{eq:A_B_C_Ns}
\end{equation}
\noindent
Finally, for $i=1$ or $i=N_{s}$ in Eq.~\ref{eq:Mdot_00_cd_sg_ox}, $A_{0}=C_{N_{s}+1}=0$ since those sections do not exist. For the last section, to accommodate the barycenter variation, a modified section spacing factor $f_{s}^{*}$ \cite{yang2019multi} is considered:
\begin{equation}
    f_{s}^{*}=\frac{1}{a}\left(\frac{\Delta V_{N_{s}-1}}{2}\right).
\end{equation}

The source terms for $M_{1,0}^{i}$ are derived from the number density source terms and given by
\begin{equation}
    \dot{M}_{1,0}^{i}=\dot{M}_{0,0}^{i}\left( I_{i} \right)V_{i}^{*},
    \label{eq:Mdot_10_cd_sg_ox}
\end{equation}
which is obtained by multiplying Eq.~\ref{eq:Mdot_00_cd_sg_ox} by $V_{i}^{*}$. Finally, the source terms for $M_{0,1}^{i}$ are
\begin{equation}
    \dot{M}_{0,1}^{i}=\dot{M}_{0,0}^{i}\left( I_{i} \frac{\delta S_i}{\delta V} \right)V_{i}^{*}.
    \label{eq:Mdot_01_cd_sg_ox}
\end{equation}
\noindent
Note that the $\dot{M}_{0,1}^{i}$ in Eq.~\ref{eq:Mdot_01_cd_sg_ox} is recast from Eq.~\ref{eq:Mdot_10_cd_sg_ox} by introducing the argument $I_{i}(\delta S_{i}/\delta V)$, similar to Mueller et al.~\cite{mueller2009hybrid}. $\delta V$ is the change of volume associated with growth or destruction processes. Then, $\delta S_{i} / \delta V$ is calculated following Eq.~\ref{eq:delta_S_def} for condensation and surface growth, whereas for oxidation \cite{blanquart2009joint} the surface area change is equal to 
\begin{equation}
    \frac{\delta S_{i}}{\delta V} = \frac{2}{3}\frac{S_{i}}{V_{i}}.
\end{equation}

The condensation, surface growth, and oxidation rates $I_{i}$ are based on detailed models from Mueller et al.~\cite{mueller2009hybrid,mueller2009joint,mueller2011modeling}. The condensation rate is written as
\begin{equation}
    I_{i}\rvert_{\rm cd}=\beta_{C_{i}} \delta V [\mathrm{DIMER}],
    \label{eq:I_cd}
\end{equation}
\noindent
where $\beta_{C_{i}}$ is the collision rate of PAH dimers and soot particles and $\delta V$ is equivalent to the volume of a PAH dimer \cite{mueller2009hybrid}. Then, the surface growth rate is given by
\begin{equation}
    I_{i}\rvert_{\rm sg}=k_{\rm sg} \chi S_{i} \delta V,
    \label{eq:I_sg}
\end{equation}
\noindent
where $k_{\rm sg}$ is surface growth rate coefficient \cite{frenklach1991detailed}, $\chi=1.7\times10^{19}\rm\:m^{-2}$ is the hydrogenated sites surface density, and $\delta V$ is equal to the volume two carbon atoms \cite{mueller2009hybrid}. 
Finally, the oxidation rate is
\begin{equation}
    I_{i}\rvert_{\rm ox}=-k_{\rm ox} \chi S_{i} \delta V,
    \label{eq:I_ox}
\end{equation}
\noindent
 where $k_{\rm ox}$ is the oxidation rate coefficient including oxidation by both ${\rm OH}$ \cite{kazakov1995detailed} and ${\rm O_{2}}$ \cite{neoh1985effect} and $\delta V$ is similar as in surface growth \cite{mueller2009hybrid}. Further details on the formulation of these terms can be found in Mueller et al.~\cite{mueller2009hybrid}.

\section{Validation in laminar flames}
\label{sec:validation_laminar}

\subsection{Experimental and computational setups}
The new BMMSM model is validated in Flame C4 from Abid et al.~\cite{abid2008evolution}, a burner-stabilized laminar premixed ethylene-oxygen-argon flame at atmospheric conditions. The inlet composition corresponds to $\phi=2.07$ ($\rm C_{2}H_{4}$: 16.3\%, $\rm O_{2}$: 23.7\%, $\rm Ar$: 60\%, by volume). The inlet velocity is $\rm 6.53~cm/s$. Computational results using the univariate MMSM are also shown to evaluate the performance of the new bivariate BMMSM. Computations where carried out in FlameMaster \cite{pitsch1998flamemaster}, with an imposed temperature profile from experimental measurements and a gas-phase kinetic mechanism consisting of 158 species and 1804 reactions, including PAHs up to four aromatic rings from Blanquart and coworkers~\cite{blanquart2009chemical,narayanaswamy2010consistent}. Simulations were performed with multiple numbers of sections, $N_{s}=6$, 8, 12, 16, and 32, in order to compare the accuracy of the models with respect to the number of sections. The section spacing factor $f_{s}$ is considered such that the ratio $V_{N_{s}}/V_{1}$ remains the same in all cases, which gives $f_{s}=333.59$, 63.43, 14.025, 6.94, and 2.56, respectively. Results presented in the following subsections are labeled according to the total number of (soot) variables solved in each case,  $N_{v}^{\rm MMSM}=2N_{s}$ and $N_{v}^{\rm BMMSM}=3N_{s}$.

\subsection{Results}

\begin{figure*}[!h]
\centering
\begin{minipage}{0.49\textwidth}
  \centering
  \includegraphics[trim={0mm 0mm 0mm 0mm},clip,width=67mm]{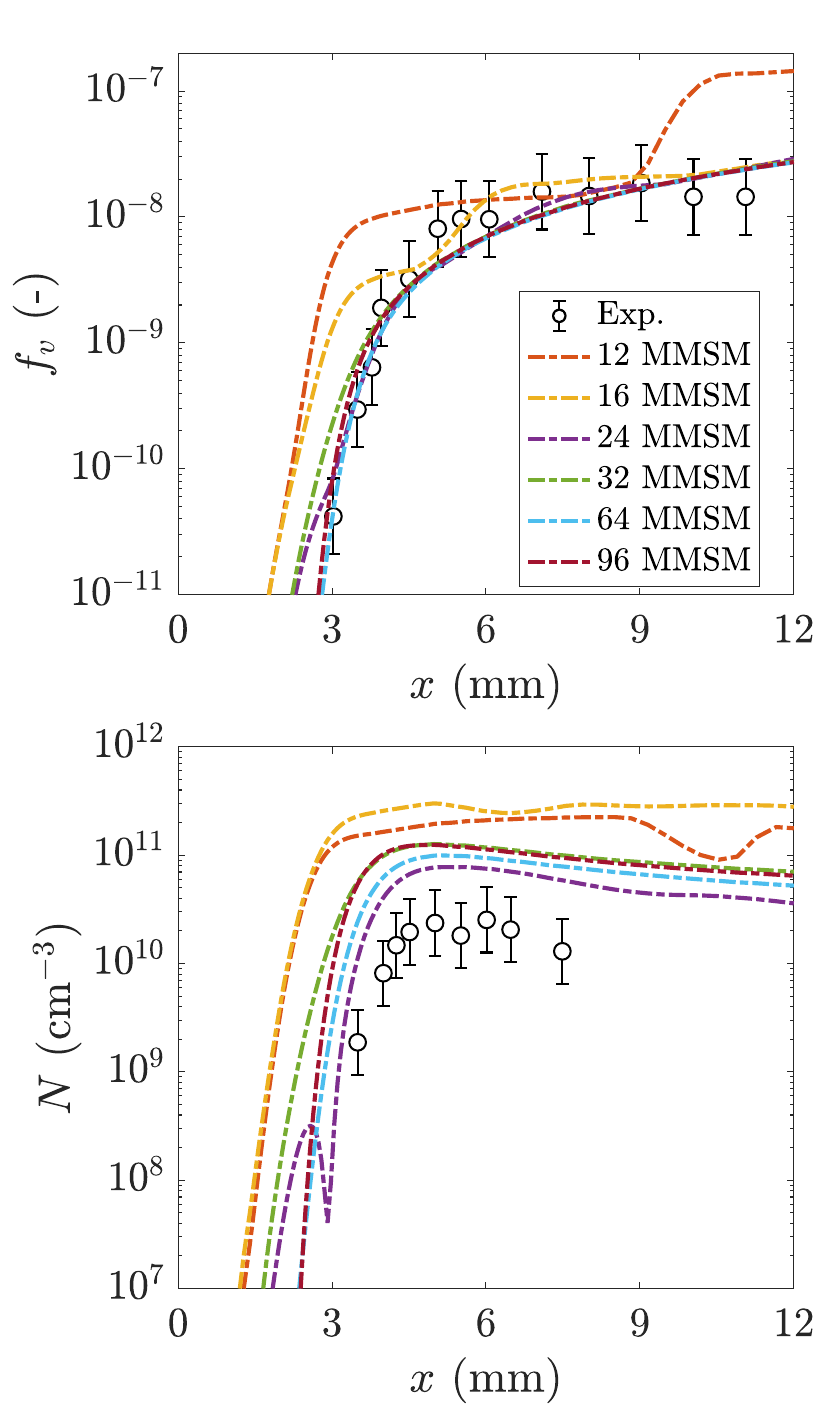}
\end{minipage}
\hspace{-0.1in}
\begin{minipage}{0.49\textwidth}
  \centering
  \includegraphics[trim={0mm 0mm 0mm 0mm},clip,width=67mm]{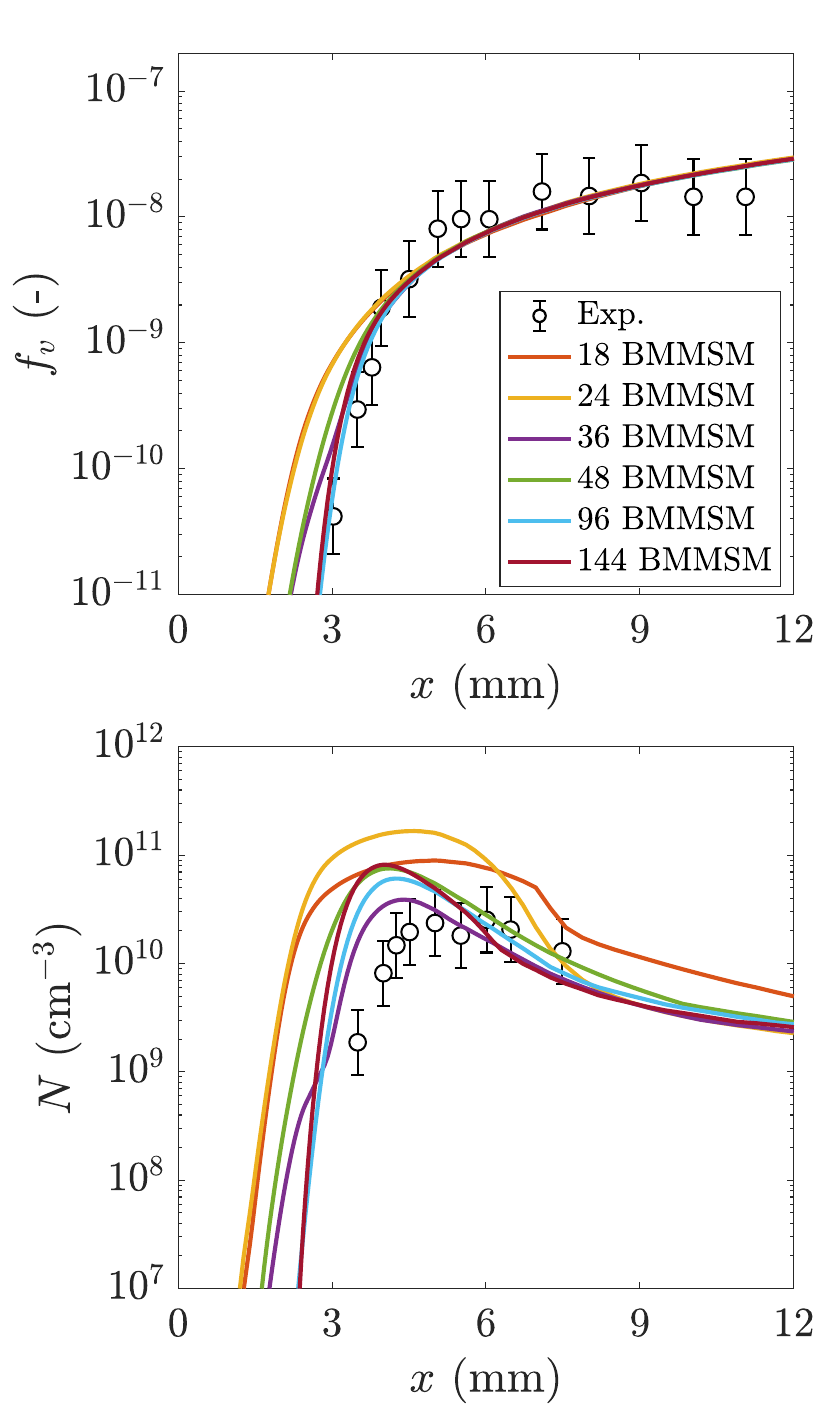}
\end{minipage}
\caption{Soot volume fraction (top) and number density (bottom) profiles calculated considering soot particles with particle diameters larger than $2.5\:\rm nm$. Computational results using the univariate MMSM (left, dash-dotted) and BMMSM (right, solid) are compared against experimental measurements \protect\cite{abid2008evolution}. The color code indicates those results using the same number of sections.}
\label{fig:fv_numdens_abid} 
\end{figure*}

Soot volume fraction and number density are calculated considering soot particles with particle diameters larger than $2.5\:\rm nm$, as specified in Ref.~\cite{abid2008evolution} due to instrument limitations. Yang and Mueller~\cite{yang2019multi} previously showed acceptable accuracy in soot volume fraction $f_{v}$ with the univariate model with as few as eight sections, which is confirmed in Fig.~\ref{fig:fv_numdens_abid} (top). With as few as six sections, BMMSM is able to predict the soot volume fraction accurately compared to both a larger number of sections and the experimental measurements. Fig.~\ref{fig:fv_numdens_abid} (bottom) shows the number density $N$ profiles. Computational results using the univariate MMSM overpredict $N$ all along the flame profile. Conversely, BMMSM improves these results by capturing the decrease of $N$ due to the bivariate characterization: for the same volume, aggregates will coagulate faster, leading to an ultimately lower number density. The variation of the soot number density with the number of sections is more significant than for the soot volume fraction, but the number density is predicted within experimental uncertainty again with as few as six sections at and beyond the peak at 4 mm.

\begin{figure*}[!h]
\centering
\includegraphics[trim={0mm 0mm 0mm 0mm},clip,height=0.78\textheight]{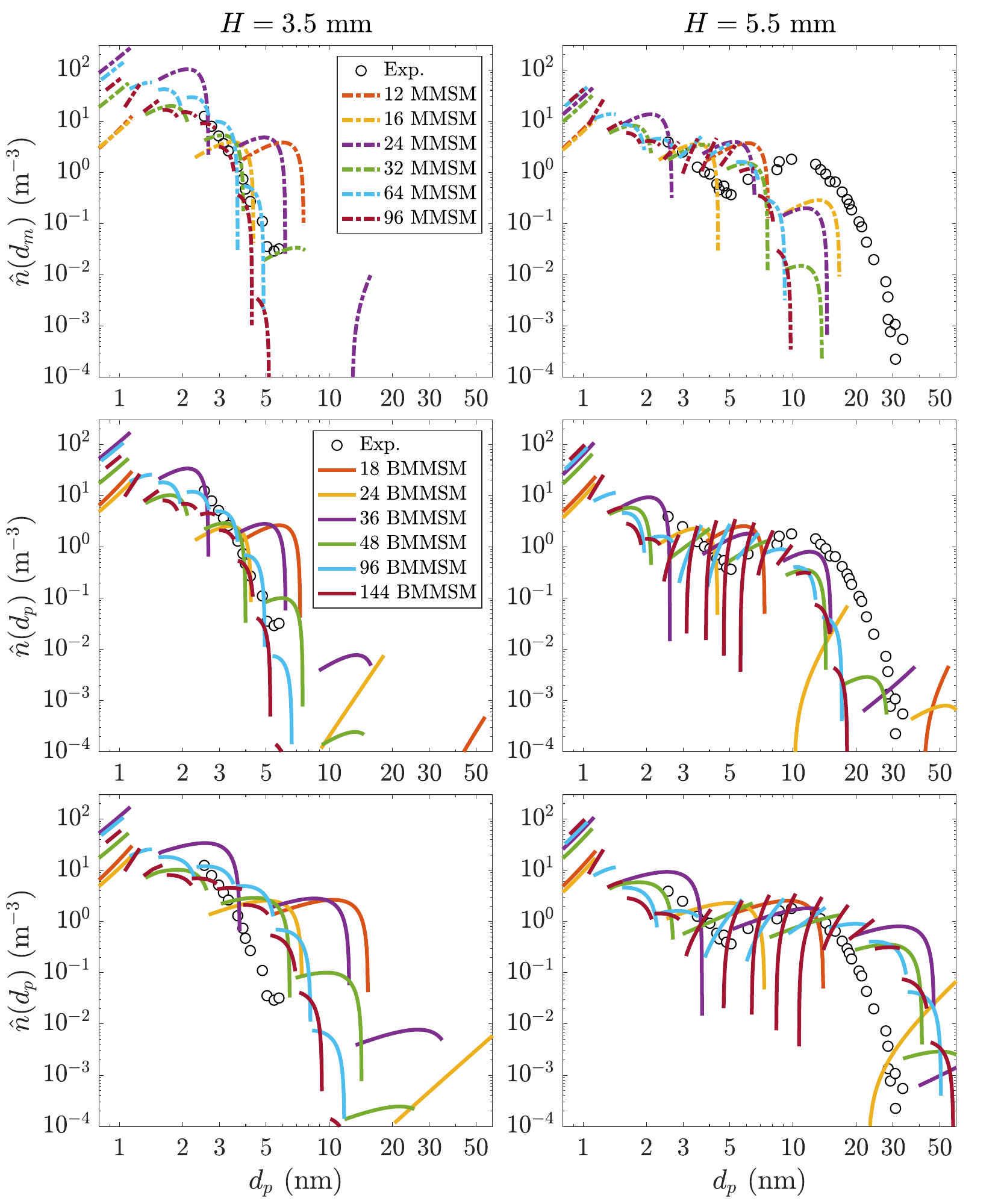}
\caption{PSDF at two locations above the burner: $H=3.5 \rm mm$ (left column) and $5.5 \rm mm$ (right column). Computational results using the univariate MMSM (dash-dotted, top row) and BMMSM (solid), considering $d_{p}$ equal to the spherical (middle) and mobility (bottom) diameters, are compared against experimental measurements (symbols) \protect\cite{abid2008evolution}. The color code indicates those results using the same number of sections.}
\label{fig:psd_3.5mm_5.5mm_abid} 
\end{figure*}

Results for the particle size distribution function (PSDF) $\hat{n}(d_{p})$ are shown in Fig.~\ref{fig:psd_3.5mm_5.5mm_abid}. The PSDF is defined as in the work of Abid et al.~\cite{abid2008evolution} as
\begin{equation}
    \hat{n}(d_{p})=\left.\frac{1}{N}\frac{{\rm d} N(d_{p})}{{\rm d} \log d_{p}}\right\rvert_{d_{p}},
    \label{eq:n_hat_def}
\end{equation}
where $d_{p}$ is the particle diameter, $N(d_{p})$ is the (reconstructed) soot particle cumulative distribution function (CDF), and $N=N(+\infty)$ is the total particle number density. Similarly to soot volume fraction and number density, the PSDF is computed by only considering soot particles with particle diameters larger than $2.5\:\rm nm$ (in the normalizing $N$)~\cite{abid2008evolution}.
Figure~\ref{fig:psd_3.5mm_5.5mm_abid} shows the PSDF at two different heights above the burner: $H=3.5~\rm mm$ (left column) and $5.5~\rm mm$ (right column). Abid et al.~\cite{abid2008evolution} observed spherical particles using a scanning mobility particle sizer (SMPS) and have defined a corrected spherical particle diameter. Therefore, BMMSM results are plotted using both the spherical diameter ($d_{p}\propto V^{1/3}$), as the univariate model, and mobility diameter ($d_{p}=d_{m}$), which is assumed equal to the collision diameter $d_{c}$ as stated in Section~\ref{sec:bmmsm}. A unimodal PSDF is obtained at $H=3.5~\rm mm$. Both univariate MMSM and BMMSM are capable of reproducing the experimental observations, with no major difference between results. At this location, the soot population is nucleation dominated, so both models predict essentially all spherical soot. A slight overprediction is observed in the bottom row using BMMSM with small $N_{s}$. Increasing $N_{s}$ improves the accuracy of results in all cases, due to improvements in predicting the initial coagulation process. A bimodal PSDF is observed at $H=5.5~\rm mm$ in the experimental measurements. For the most part, MMSM still predicts a unimodal distribution and a maximum size far smaller than the experimental measurements. Conversely, BMMSM correctly predicts both a bimodal size distribution with the presence of much larger particles, with a small deviation between results using different $d_{p}$ definitions. Unsurprisingly, the bivariate description of soot is required to accurately reproduce the evolution of the size distribution. In all cases, the PSDF within each section present a parabolic shape, which is due to the linear approximation of the NDF in volume as indicated in Eq.~\ref{eq:nV_approx} (for $i<N_{s}$). Indeed, the soot volume $V$ is proportional to the cube of the particle diameter $V \propto d_{p}^3$, so the derivative in Eq.~\ref{eq:n_hat_def} in terms of volume is proportional to $V^2$.



\section{Application to turbulent sooting flames}
\subsection{LES modeling framework}

In this work, the Large Eddy Simulation (LES) modeling framework includes the aforementioned soot model, a combustion model, and a turbulence-chemistry-soot interactions model. The latter two models are succinctly described below with complete details provided in the cited works.

The Radiation Flamelet/Progress varable (RFPV) with soot considerations \cite{mueller2012model,yang2019large} is utilized to model the combustion. This approach characterizes the thermochemical state by parameterizing it in terms of the mixture fraction ($Z$), progress variable ($C$), and heat loss parameter ($H$), which are obtained from solving the nonpremixed manifold equations in mixture fraction (steady flamelet equations). To account for soot formation, the mixture fraction has a compensatory source term $\dot{m}_{Z}$ for the local mixture leaning caused by the removal of PAHs from the gas phase. Similarly, the source term for the progress variable is adjusted to consider the local variation in effective fuel resulting from the removal of PAHs \cite{mueller2012model}. The heat loss parameter source term captures radiative heat losses with $H = 0$ corresponding to the adiabatic state. The radiation model employs an optically thin gray approach, including gas effects, as in Barlow et al.~\cite{barlow2001scalar}, and soot effects, as in Hubbard and Tien~\cite{hubbard1978infrared}. A Strain-Sensitive Transport Approach (SSTA) is utilized to manage different effective Lewis numbers of species, based on their characteristic length scales \cite{yang2020large}. In addition, a lumped PAH transport equation is solved to address their slower chemistry compared to other combustion products in the thermochemical database, following the approach developed by Mueller and Pitsch \cite{mueller2012model}. The lumped PAH source term is separated into contributions from chemical production, chemical consumption, and dimerization.

Turbulence-chemistry-soot interactions require closure at the subfilter scales. To close turbulence-chemistry subfilter interactions, convolution is performed for each manifold solution in the database with a presumed beta subfilter Probability Density Function (PDF) for the mixture fraction. The resulting thermochemical database is then stored in a lookup table, in terms of the filtered mixture fraction $\widetilde{Z}$, subfilter mixture fraction variance $Z_{v}$, filtered progress variable $\widetilde{C}$, and filtered heat loss parameter $\widetilde{H}$. To close turbulence-soot subfilter interactions, a presumed soot subfilter PDF model that captures finite-rate oxidation of soot is employed, following the work of Maldonado Colmán et al. \cite{hmcolman2023from}, which was validated for turbulent nonpremixed jet flames \cite{hmcolman2023from,duvvuri2023relative} and bluff body flames \cite{hmcolman2023bbf}. This model is based on the presumed bimodal PDF of Mueller and Pitsch~\cite{mueller2011modeling}, which considered sooting and non-sooting modes. 
The sooting mode profile accounts for a transition from rich to lean mixtures to account for the oxidation of soot as it encounters rich mixtures. Previous work of Yang et al.~\cite{yang2019large} considered an abrupt transition as soon as the soot oxidation rate surpassed the soot surface growth rate, which assumes that soot oxidation is strictly mixing controlled so oxidation infinitely fast. In the work of Maldonado Colmán et al.~\cite{hmcolman2023from}, which is adapted to BMMSM here, this abrupt transition was replaced with a smooth transition to account for finite-rate soot oxidation compared to the local soot transport rate, which depends on the local mixture fraction dissipation rate. The subfilter intermittency, that is, the weight between the sooting and non-sooting modes, is obtained by solving an additional transport equation for $N^{2}$, where $N=\sum_i M_{0,0}^{i}$, which is similar to HMOM \cite{mueller2011large,mueller2012model}.


\subsection{Experimental and computational details}
BMMSM is evaluated with the KAUST turbulent nonpremixed jet flame, investigated experimentally by Boyette and coworkers~\cite{boyette2017soot,chowdhury2017time}. The experimental configuration is similar to the Sandia sooting flame \cite{zhang2011design} but with a nitrogen-diluted central jet ($\rm C_{2}H_{4}:~35\%$, $\rm N_{2}:~65\%$, by volume). The inner jet diameter is $D=3.2~\rm mm$ and bulk velocity $54.7~\rm m/s$, with a $\rm Re=20,000$. Both pilot flame (ethylene-air mixture with $\phi=0.9$) and surrounding air coflow are kept the same as the original configuration. Further details about this burner can be found in Refs.~\cite{boyette2017soot,chowdhury2017time,zhang2011design}. Experimental measurements of the particle size distribution function were obtained with a SMPS in Refs.~\cite{boyette2017soot,chowdhury2017time} at several positions along the flame centerline.

LES computations were carried out using the NGA structured finite difference solver for low  Mach number turbulent reacting flows \cite{desjardins2008high, macart2016semi}. The grid-filtered LES equations are computed in a similar domain as in Maldonado Colmán et al.~\cite{hmcolman2023from} for the Sandia sooting flame, with dimensions of $300D\times75D$ in the axial and radial directions, respectively, and discretized with $192\times96\times32$ grid points in the axial, radial, and circumferential directions, respectively. Inlet boundary conditions are prescribed as in Ref.~\cite{hmcolman2023from}, precomputing the unsteady velocity inflow of the central jet following the experimental conditions and the coflow velocity set to $0.6~\rm m/s$. Only the central jet mixture composition is modified to match the current case. Lagrangian dynamic Smagorinsky(-like) models \cite{germano1991dynamic,meneveau1996lagrangian} are considered to close the subfilter stress and scalar flux terms. The kinetic mechanism for the gas-phase is the same as the one from Section~\ref{sec:validation_laminar} \cite{blanquart2009chemical,narayanaswamy2010consistent}.

Two simulations are performed with BMMSM with 8 and 12 sections, resulting in 24 and 36 unknowns, with spacing factors $f_{s}=8.8327$ and 4.0, respectively, in order to focus in the experimental soot size region (2 to 70 $\rm nm $ \cite{boyette2017soot}). A simulation using the Hybrid Method of Moments (HMOM) \cite{mueller2009hybrid} is also performed in order to compare BMMSM's performance. In all cases, the presumed subfilter PDF model for finite-rate oxidation of soot from Ref.~\cite{hmcolman2023from} is considered for subfilter turbulence-chemistry-soot interactions as described in the previous section. The total duration of simulation is 150 ms, which is equivalent to approximately 5 flow-through times through along the sooting region of the flame ($x/D\approx140$), which is sufficient for statistical convergence.

\subsection{Temperature, soot volume fraction, and total number density}

\begin{figure}[!t]
\centering
\includegraphics[trim={0mm 0mm 0mm 0mm},clip,width=72 mm]{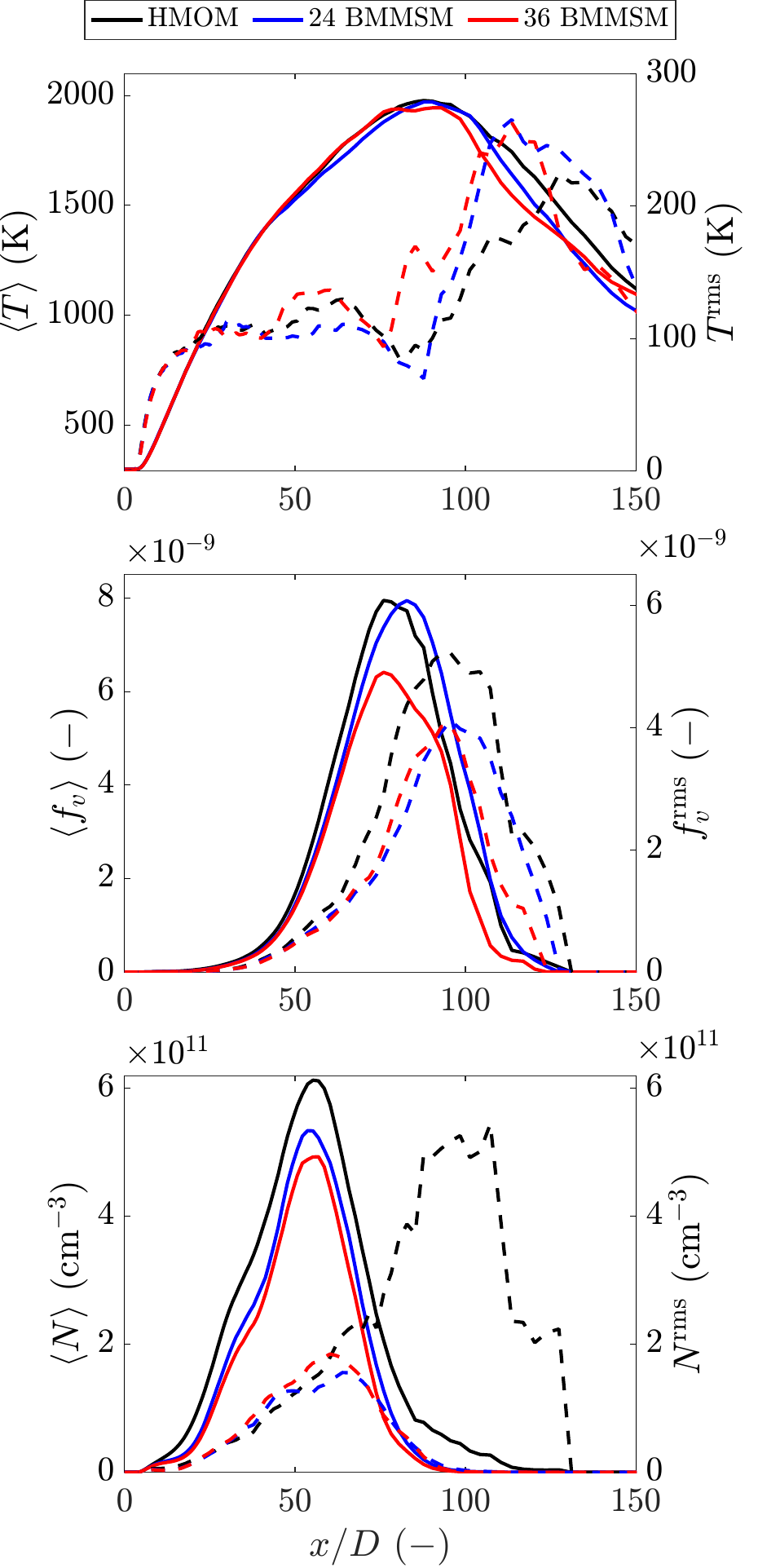}
\caption{Centerline profiles of mean (left axes, solid lines) and rms (right axes, dashed lines) temperature (top), soot volume fraction (center), and total number density (bottom). Computational results using the BMMSM with 8 and 12 sections and HMOM are compared.}
\label{fig:T_fv_kaust}
\end{figure}

Figure~\ref{fig:T_fv_kaust} shows centerline profiles of the mean (continuous) and rms resolved fluctuations (dashed) of the temperature (top), soot volume fraction $f_{v}$ (center), and total number density $N$ (bottom). Although no experimental measurements exist for these three quantities, computational results are shown using HMOM and using BMMSM with 8 and 12 sections, with the aim to assess the influence of soot statistical approach and number of sections. The mean and rms temperature profiles show a good agreement between models with a slight difference after the temperature peak. Note the significant increase in the relative magnitude of the rms temperature fluctuations in the downstream portion of the flame, which corresponds to the maximum in the rms soot volume fraction fluctuations so likely due to soot radiation fluctuations.

The $f_{v}$ profiles show that BMMSM predicts the same or lower soot volume fraction compared to HMOM, which is favorable because this modeling framework with HMOM has been shown to overpredict the volume fraction in such configurations \cite{hmcolman2023from,duvvuri2023relative}. For both the mean soot volume fraction and the mean soot number density, the BMMSM results with fewer sections are comparable to HMOM, which is consistent with prior work in laminar flames \cite{yang2019multi}. While the maximum mean soot volume fraction in BMMSM is somewhat sensitive to the number of sections (about 1.4 ppb difference between 8 and 12 sections), the results are not strongly sensitive. The soot volume fraction fluctuation profiles indicate that the number of sections does not have as strong of an influence as the soot model, with differences between HMOM and BMMSM of about 2 ppb in the peak. The mean number density profiles using BMMSM show little variation with respect to the number of sections and are lower compared to using HMOM, whose peak value is about $20\%$ higher. The most significant qualitative difference between HMOM and BMMSM is observed in the oxidation region of the flame, where HMOM predicts significant number density for regions up to $x/D\approx115$. Here, the $N$ rms fluctuations are much greater using HMOM compared to using BMMSM. Clearly, HMOM and BMMSM are predicting fundamentally different soot oxidation processes. The explanation for these trends in the number density findings are further analyzed later.

\subsection{Soot intermittency and soot temperature}

\begin{figure}[!h]
\centering
\includegraphics[trim={0mm 0mm 0mm 0mm},clip,width=67 mm]{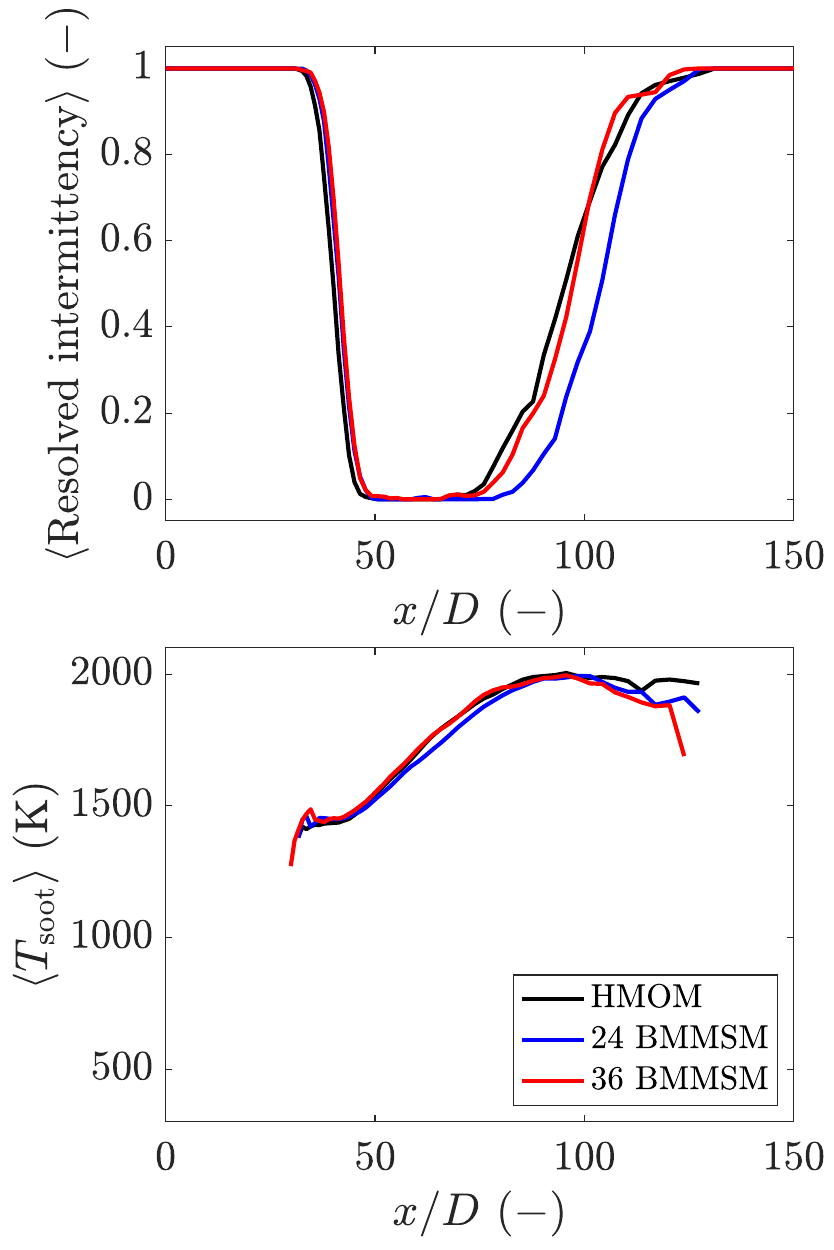}
\caption{Centerline profiles of resolved intermittency (top) and mean soot temperature (bottom). Computational results using the BMMSM with 8 and 12 sections and HMOM are compared.}
\label{fig:rint_Ts_kaust}
\end{figure}

To understand better the nature of soot fluctuations in the previous subsection, computational results of resolved soot intermittency and soot temperature ($T_{\soot}$) are analyzed. The resolved soot intermittency denotes the probability of not finding soot at a spatiotemporal location conditioned on a threshold criterion of soot volume fraction, which is usually established by the experimentalist. Essentially, the soot intermittency is obtained by time-averaging binary detection of soot, where unity indicates that no soot was present over the threshold. Since no experimental measurements of soot intermittency or volume fraction are available for this configuration, the threshold is set at 3\% of the maximum value of soot volume fraction observed in simulations, i.e., about 0.5 ppb, which was utilized in recent experiments by Boyette and coworkers~\cite{boyette2021effects}. Figure~\ref{fig:rint_Ts_kaust} (top) shows computational results of soot intermittency using BMMSM and HMOM. The sooting region is well defined between the abrupt intermittency drop at about $x/D=40$ in the soot growth region, consistent with low rms soot volume fraction in Fig.~\ref{fig:T_fv_kaust}, and by a more moderate increase downstream starting at $x/D=75$ in the oxidation region, where larger fluctuations in the soot volume fraction are observed in Fig.~\ref{fig:T_fv_kaust}. Interestingly, the differences in the intermittency more or less mirror the differences in the soot volume fraction profiles, with only small differences in the oxidation region. Therefore, the differences in the oxidation region in the number density and its fluctuation are due to inherent differences in the soot statistical models and how particles of different sizes interact with turbulence rather than some fundamental difference in the overall sooting flame dynamics.

Mean soot temperature profiles are plotted in Fig.~\ref{fig:rint_Ts_kaust} (bottom), which is computed by locally averaging temperature values when the instantaneous resolved soot intermittency is zero. BMMSM profiles are similar to that of HMOM with differences only appearing once soot has nearly disappeared. The influence of number of sections on soot temperature is negligible. This means that the location of soot with respect to mixture fraction is essentially the same between the statistical models and is much more sensitive to the subfilter soot-turbulence interactions model \cite{hmcolman2023from}, which is common between BMMSM and HMOM.

\subsection{Size distribution}

\begin{figure*}[!h]
\centering
\includegraphics[trim={0mm 0mm 0mm 5mm},clip,height=0.84\textheight]{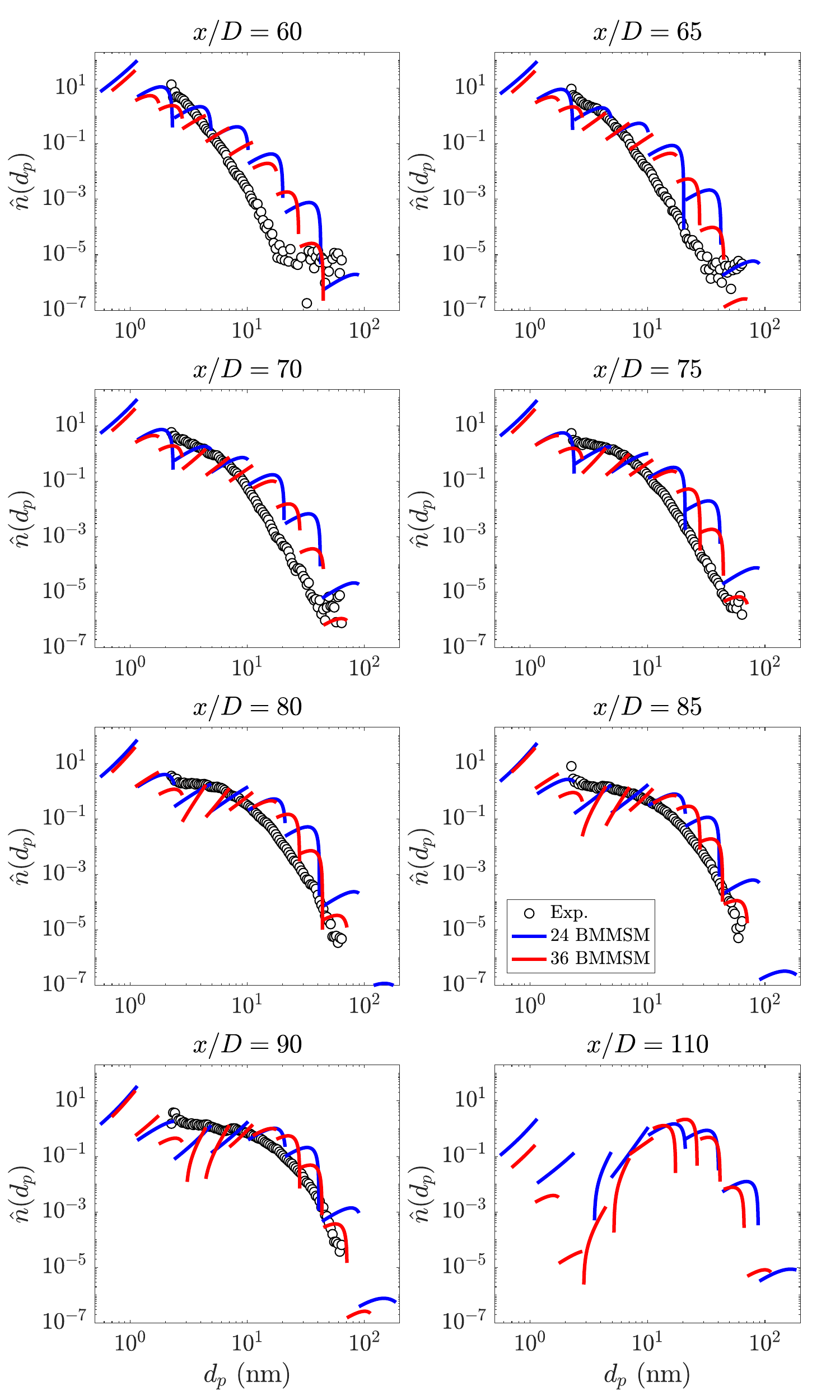}
\caption{PSDF at several locations along the centerline depicting the progress from a unimodal to a bimodal distribution. Computational results using BMMSM with 8 and 12 sections (lines) are compared against experimental measurements (symbols) {\protect\cite{boyette2017soot}}.}
\label{fig:PSDF_kaust_CL8}
\end{figure*}

The computational results of the PSDF using BMMSM were obtained at several locations downstream of the burner along the centerline from $x/D=60$ to 90, with $\Delta x/D=5$, and an eighth one further downstream at $x/D=110$ and are plotted in Fig.~\ref{fig:PSDF_kaust_CL8}. Experimental data are available only for the first seven locations \cite{boyette2017soot}, including only soot particles with particle diameter larger than $2\rm\:nm$ and smaller than $225\rm\:nm$. The PSDF is calculated following Eq.~\ref{eq:n_hat_def}, which is normalized using the total particle number
density within the experimental detection limits. The PSDF evolution is well captured: a small amount of large soot particles is found closer to the burner, which effectively grows as they travel further downstream from the burner. The different plots suggest that BMMSM is capable of reproducing the transition from unimodal PSDF in the first seven locations, which is supported by the experimental data, to bimodal PSDF in the last location. It is also noticeable that the predicted PSDF is not a strong function of the number of sections. Overall, BMMSM is very accurate along the entire flame compared to measurements. Nevertheless, the model exhibits a slight overprediction of the PSDF in the mid-size region ($d_{p}\simeq 20\rm\:nm$) near the burner ($x/D\leq70$), which diminishes as it progresses downstream. Moreover, BMMSM results are in line with other results in the literature \cite{schiener2019transported} but requiring fewer degrees of freedom. In the present study, the use of the BMMSM method reduces the required number of soot scalars to 24 or 36 for fractal aggregate soot, whereas the sectional model from Ref.~\cite{schiener2019transported} used 62 soot scalars for only spherical soot.

\begin{figure*}[!t]
\centering
\includegraphics[trim={0mm 0mm 0mm 0mm},clip,width=120 mm]{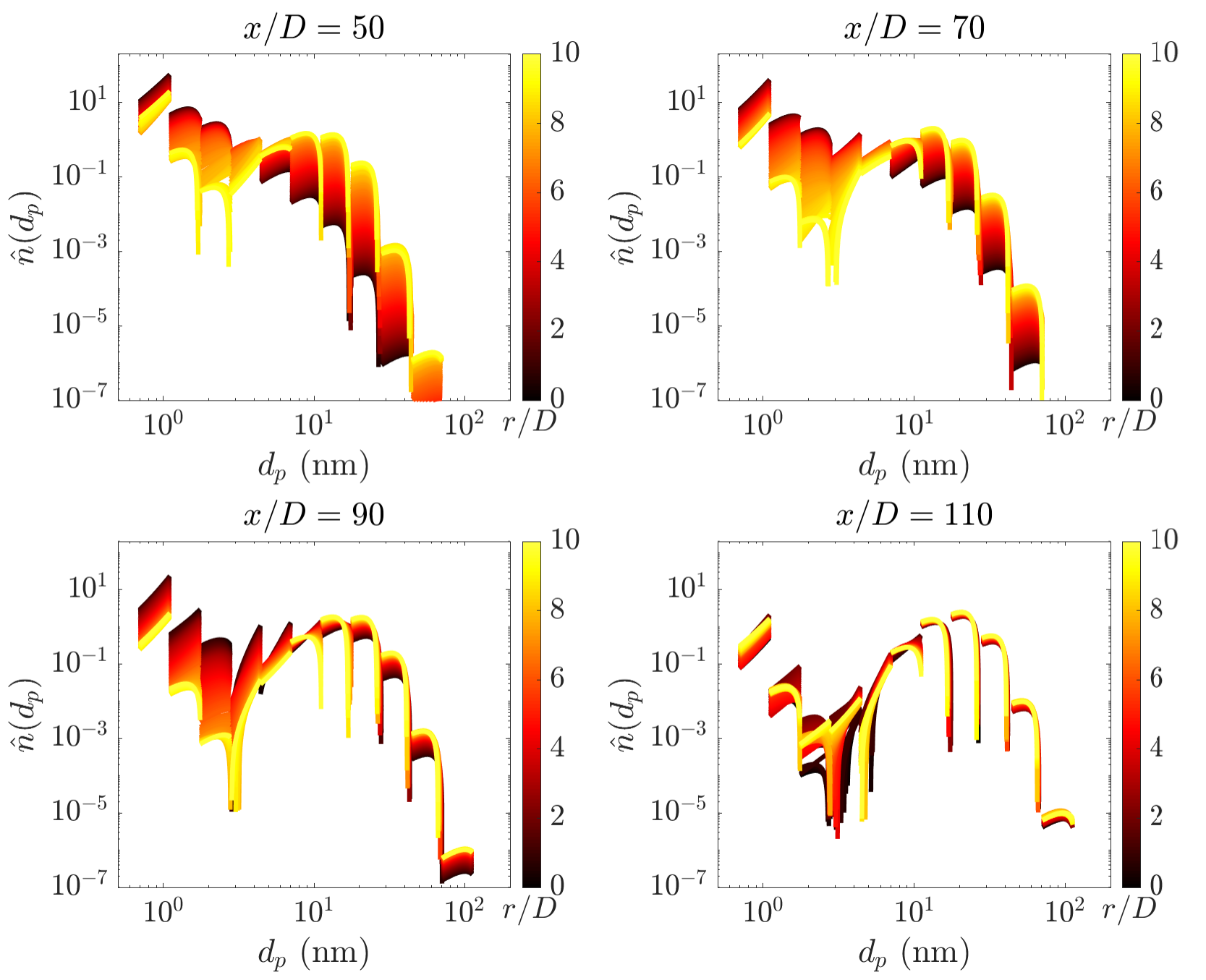}
\caption{Variation of the PSDF in the radial direction (radial distance indicated by the colorbar) at four different locations downstream of the burner. Computational results using BMMSM with 12 sections are shown.}
\label{fig:PSDF_kaust_rad4}
\end{figure*}

Since experimental measurements revealed the evolution of the PSDF in the streamwise direction only, additional computational analysis is conducted to assess the radial variation of the PSDF at different locations downstream of the burner. Figure~\ref{fig:PSDF_kaust_rad4} shows the radial evolution of the PSDF at four locations downstream of the burner: $x/D=50$, 70,  90, and 110. Each location is representative of a portion of the sooting region (see Fig.~\ref{fig:rint_Ts_kaust}). The colorbar represents the normalized radial distance ($r/D$) of the PSDF, with darker colors being nearer the centerline. For particles with small sizes ($d_{p}\leq30\rm:nm$), the PSDF decreases radially outward in the first three locations, which shifts from nucleation-dominated to growth-dominated moving away from the centerline. In the last location, where oxidation dominates, the PSDF remains nearly constant. Furthermore, in the first three locations, moving away from the centerline, a ``trough'' emerges in the PSDF for particle diameters ranging from 20 to $30\rm\:nm$, indicating the formation of the second mode. At $x/D=110$, the already existent trough is slightly shifted to smaller diameters, which is attributed to the oxidation process, whereas the magnitude of PSDF is practically unchanged or increases (as the total number density decreases). For particles with larger sizes ($d_{p}>30\rm\:nm$), the presence of a second mode is evident already near the burner at large radial distances and becomes more apparent closer to the centerline with increasing downstream distance. Basically, the core of the flame near the centerline is more unimodal while moving toward the periphery of the sooting region of the flame results in a bimodal distribution. However, in the last location, the second mode remains practically unchanged, with a slight shift towards smaller particle diameters.

\subsection{Source terms}
\begin{figure*}[!t]
\centering
\begin{minipage}{\textwidth}
  \centering
  \includegraphics[trim={0mm 0mm 0mm 0mm},clip,width=\textwidth]{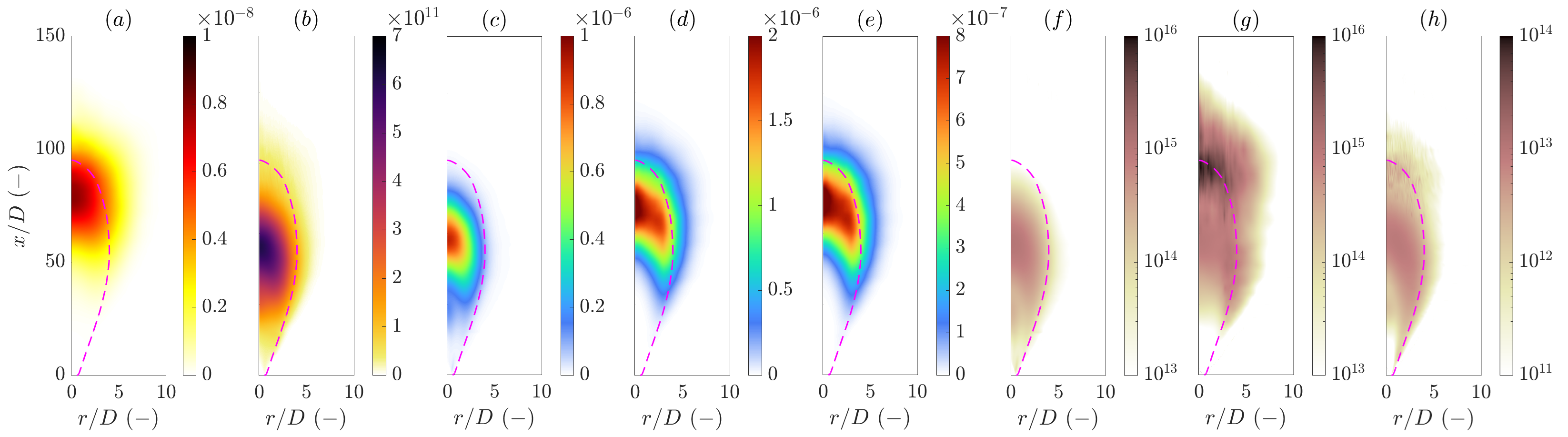}
\end{minipage}  
\begin{minipage}{\textwidth}
  \centering
  \includegraphics[trim={0mm 0mm 0mm 0mm},clip,width=\textwidth]{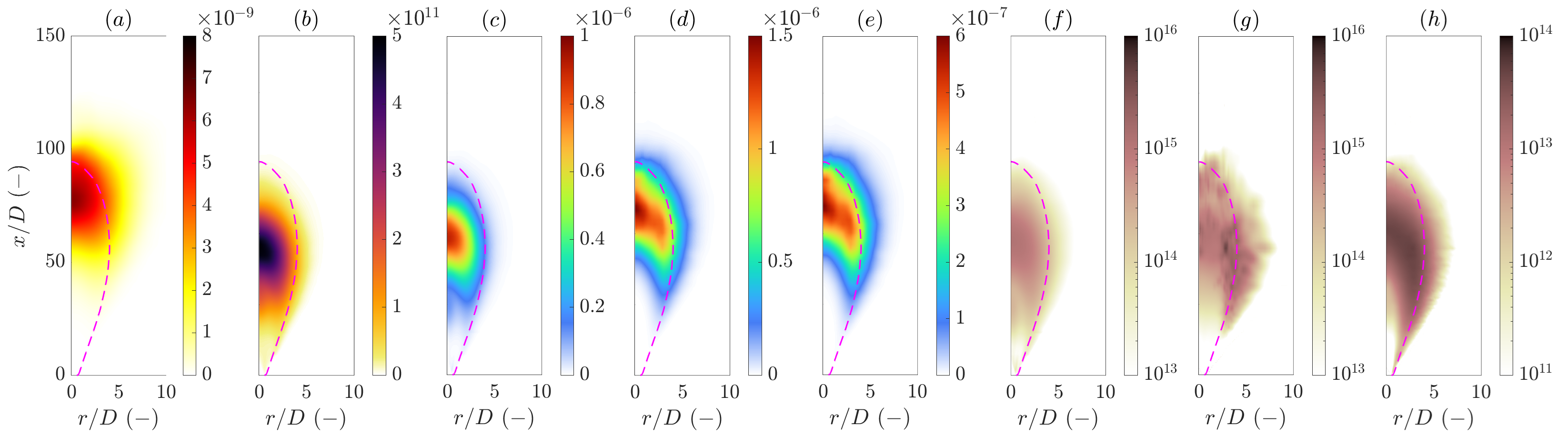}
\end{minipage}
\caption{Fields of mean soot-related quantities by columns: $(a)$ soot volume fraction $f_{v}$; $(b)$ total number density $N$; soot volume fraction source terms ${\rm d}f_{v}/{\rm d}t\rm\:[s^{-1}]$ for $(c)$ nucleation and condensation, $(d)$ surface growth, and $(e)$ oxidation (magnitude); and total number density source terms ${\rm d}N/{\rm d}t\rm\:[m^{-3}s^{-1}]$ for $(f)$ nucleation, $(g)$ coagulation (magnitude), and $(h)$ oxidation  (magnitude). Computational results using HMOM (top row) and BMMSM (12 sections, bottom row) are compared. The dotted line corresponds the mean stoichiometric mixture fraction iso-contour.}
\label{fig:st_kaust} 
\end{figure*}

A comprehensive examination of the soot source terms within the flame enables a more refined understanding of the mechanism of evolution of the soot particle size distribution. Mean soot source term fields are evaluated from simulations using BMMSM with 12 sections and are compared to those using HMOM. The findings obtained with 8 sections closely align with those obtained with 12 sections, leading to similar conclusions. Figure~\ref{fig:st_kaust} shows results using HMOM (top row) and BMMSM (bottom row), and the columns correspond to fields of mean soot-related quantities: soot volume fraction ($f_{v}$), total number density ($N$),  and their source terms (${\rm d}f_{v}/{\rm d}t$ and ${\rm d}N/{\rm d}t$, respectively). The magenta dashed line indicates the mean stoichiometric mixture fraction iso-contour $\langle Z \rangle=Z_{\stoi}$, which represents the mean position of the flame front.

The fields of soot volume fraction in column $(a)$ show good agreement between BMMSM and HMOM, with a 20\% disparity in the maximum values consistent with the centerline profiles discussed above. The peak is located in the mean rich region, as indicated by the $Z_{\stoi}$ isocontour. However, in column $(b)$, HMOM overestimates the total number density compared to BMMSM, showing a difference of about 40\%, again consistent with the centerline profiles discussed above. Additionally, the number density demonstrates a much more significant presence of soot in regions where the mean mixture faction is fuel-lean in HMOM compared to BMMSM, even with the same location of the maximum.

Column $(c)$ shows the source term of the soot volume fraction resulting from the combined effect of nucleation and condensation. Very similar results are obtained for both HMOM and BMMSM, which occurs primarily in the rich region, since this a function only of the gas-phase soot precursors. On the other hand, the surface growth (column $(d)$) and oxidation (column $(e)$) source terms exhibit higher rates of soot volume fraction production when using HMOM. However, the explanation for this trend is simple: the soot volume fraction is higher in HMOM compared to BMMSM, and the surface growth and oxidation rates scales with the amount of soot so are expected to be larger in magnitude.

Column $(f)$ shows the field of the rate of change of total soot number density due to nucleation. Both HMOM and BMMSM show similar results, with BMMSM results radially spread further than HMOM; this minor difference is simply due to lower number density predicted by BMMSM in regions closer to the mean $Z_\stoi$ resulting in less condensation relative to nucleation. Fields of coagulation in column $(g)$ show more significant differences between models. HMOM predicts the peak coagulation rate at the mean $Z_\stoi$ contour along the centerline, while BMMSM predicts the peak coagulation also along the $Z_\stoi$ contour but closer to the burner and away from the centerline. This explains the increased number density in HMOM compared to BMMSM along with the centerline with a delay in coagulation until further downstream. Likewise, the fields of column $(h)$ for oxidation rates also exhibit qualitative differences. The magnitude of the oxidation rate for the number density is much higher for BMMSM (despite the volume fraction magnitude being lower) compared to HMOM. Clearly, even though the mean soot volume fraction is comparable between HMOM and BMMSM, the two approaches provide fundamentally different evolutions of the total number density so, by extension, also mean particle size. The reason for this differences can be explained by analyzing individual sections.

\begin{figure*}[!t]
\centering
\centering
\begin{minipage}{\textwidth}
  \centering
  \includegraphics[trim={0mm 10mm 0mm 0mm},clip,width=\textwidth]{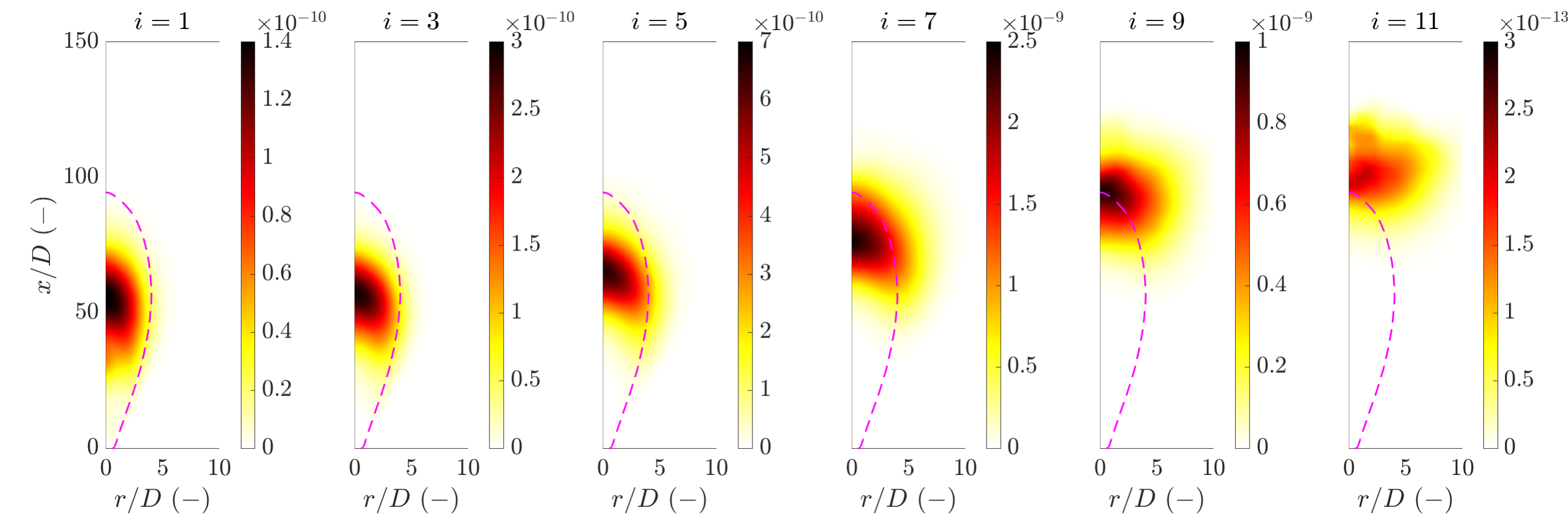}
\end{minipage}  
\begin{minipage}{\textwidth}
  \centering
  \includegraphics[trim={0mm 0mm 0mm 0mm},clip,width=\textwidth]{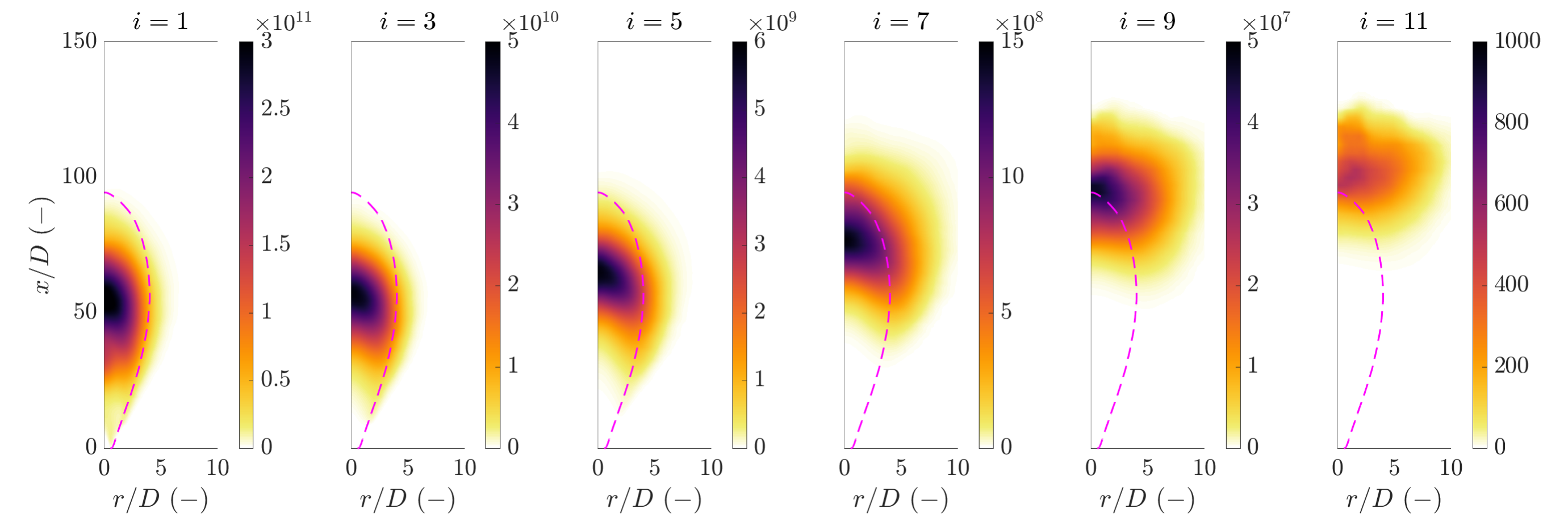}
\end{minipage}
\caption{Fields of mean soot volume fraction (top) and number density (bottom): contribution by odd sections $i=1$ to 11. Computational results using BMMSM (12 sections) are shown. The dotted line corresponds the mean stoichiometric mixture fraction iso-contour.}
\label{fig:qsoot_per_section_BMMSM36}
\end{figure*}

The evolution of the contribution of the odd sections $i=1$ to 11 on the soot volume fraction is shown in Fig.~\ref{fig:qsoot_per_section_BMMSM36} (top). Particles of larger size are located preferentially downstream. A significant decrease of soot volume fraction is observed in the latest sections, which indicates few big particles exists and that the choice of the spacing factor was appropriate. Similarly, the fields of number density by section are plotted in Fig.~\ref{fig:qsoot_per_section_BMMSM36} (bottom).  Although the distribution by section looks qualitatively similar as the soot volume fraction, quantitatively the number density fields maximizes in the first section and decreases as the local section number increases. By focusing on specific axial and radial locations, bimodality can be identified. For instance, at $x/D=90$ and $r/D=2.5$, significant soot is formed from section $i=1$ to 9 and peaks between sections $i=5$ to 7. Similar behavior is observed further downstream, and the peak value moves toward larger sections. This is analogous to what it was observed in Fig.~\ref{fig:PSDF_kaust_rad4}.

\begin{figure*}[!h]
\centering
\centering
\begin{minipage}{\textwidth}
  \centering
  \includegraphics[trim={0mm 10mm 0mm 0mm},clip,width=\textwidth]{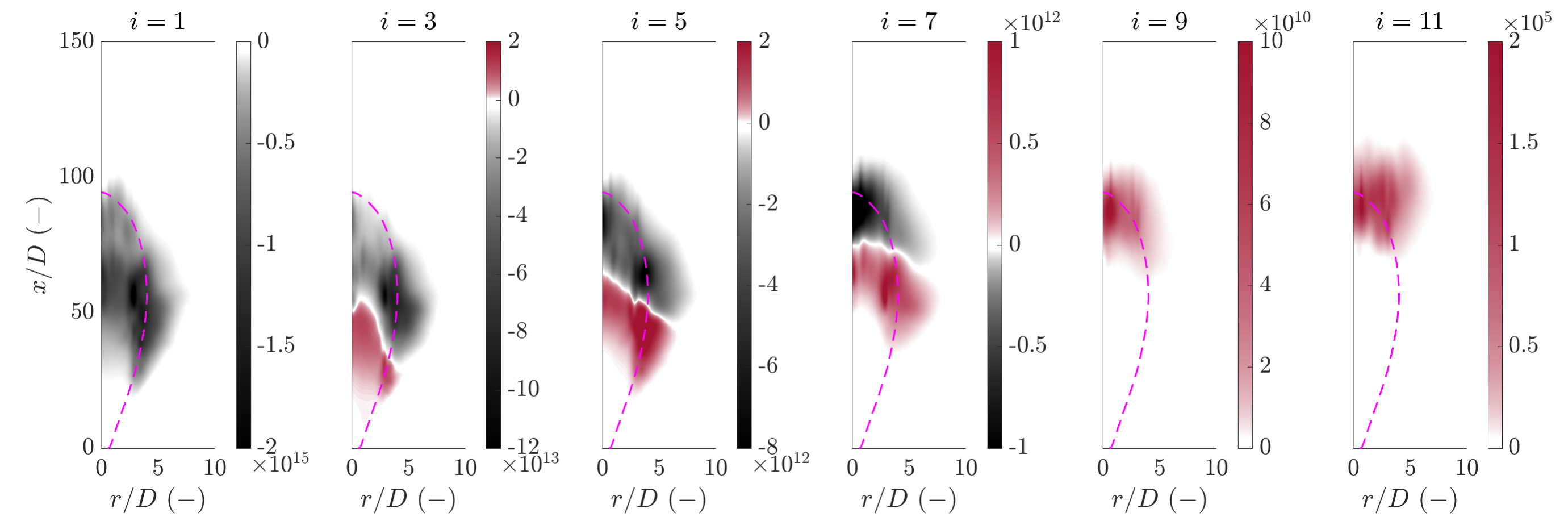}
\end{minipage}  
\begin{minipage}{\textwidth}
  \centering
  \includegraphics[trim={0mm 0mm 0mm 0mm},clip,width=\textwidth]{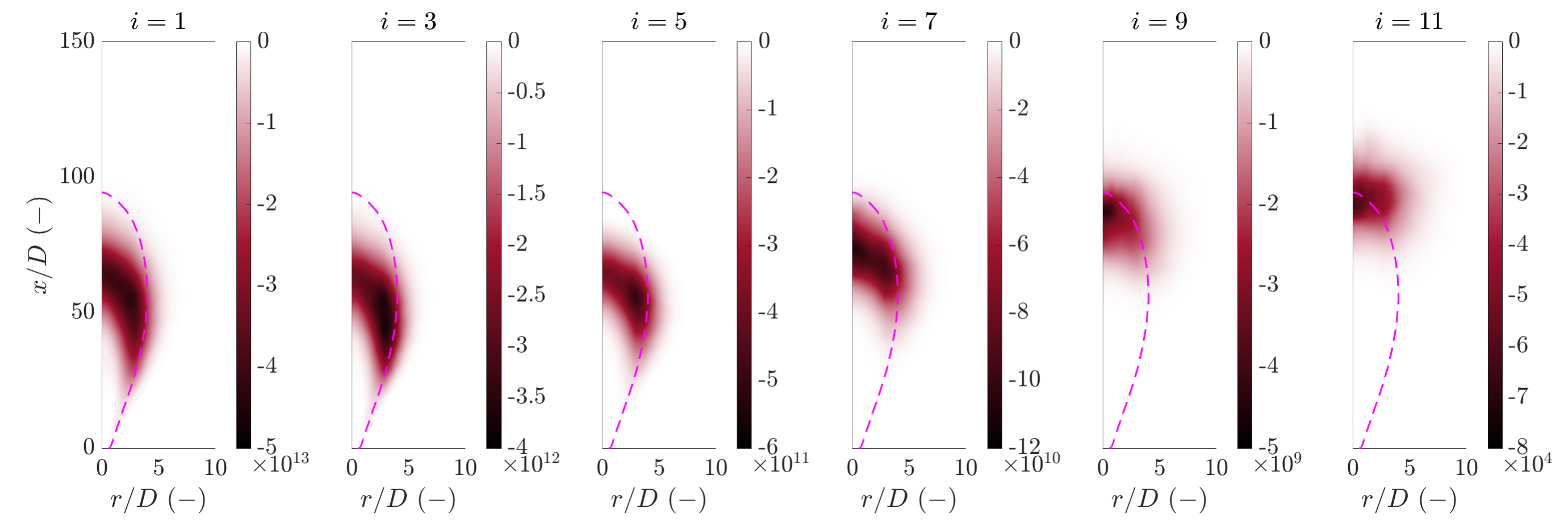}
\end{minipage}
\caption{Fields of the rate of total soot number density change due to coagulation (top) and oxidation (bottom): contribution by odd sections $i=1$ to 11. Computational results using BMMSM (12 sections) are shown. The dotted line corresponds the mean stoichiometric mixture fraction iso-contour.}
\label{fig:Nsoot_st_per_section_BMMSM36}
\end{figure*}

To understand better the behavior of number density source terms, observed in columns $(g)$ and $(h)$ of Fig.~\ref{fig:st_kaust}, the fields of the contributions by odd sections $i=1$ to 11 for coagulation (top) and oxidation (bottom) are plotted in Fig.~\ref{fig:Nsoot_st_per_section_BMMSM36}. The rates due to coagulation indicate a transition from negative values to positive values as the local section number increases: the smallest particles are only lost to form larger particles and the largest particles predominantly formed from smaller, with the small-large transition increasing in size with downstream distance as larger and larger particles are formed. The coagulation and oxidation rates of small particles are greater in magnitude than those of large particles, owing to their higher number density. Larger particles tend to oxidize closer to the mean $Z_{\stoi}$ contour since this is where the particles are located. Note that the coagulation dynamics governing the particle sizes by downstream distance: the distribution of smaller particle decreases as the coagulation rate turns negative as downstream distance increases. On the other hand, the distribution of larger particles increases as the coagulation rate tends to remain positive in the further downstream distance, only decreasing in magnitude as for the largest particles until oxidation dominates and destroy the remaining soot particles. The oxidation rate is then a reflection of where the particles are and the dependence of rate on size. Basically, the smaller particles, which are located further upstream, are oxidized quickly and disappear, while the larger particles, which are located further downstream, are oxidized more slowly so reach further downstream distances. This is directly correlated to the observations in Fig.~\ref{fig:PSDF_kaust_rad4}. The first mode peak decreases in both streamwise (by coagulation) and radial (both coagulation and oxidation) directions. The second mode peak increases in the streamwise direction more than in the radial direction due to the competition between coagulation and oxidation as the particles reach the flame front and leaner regions. Overall, BMMSM gives a larger loss of particles since first-order HMOM only has one large particle size rather than a distribution of large particle sizes so cannot capture this size-dependent oxidation process.


\subsection{Computational costs}
The computational costs of simulations using BMMSM and HMOM are assessed in Table~\ref{tab:computational_costs}. The computational costs, expressed in time per time step, are averaged over 10,000 time steps using the same computational resources. Specific costs to both solve the scalar transport equations and evaluate the soot source terms are also examined. Remarkably, on average, BMMSM simulations cost about 1.44 and 1.86 times more than HMOM in total using 8 and 12 sections, respectively. Within these, the combined costs of scalar transport and soot calculations are about 2 and 3.2 times more than HMOM, respectively. Additionally, by varying the computational resources for each simulation (i.e., more or less compute nodes), the cost of BMMSM compared to HMOM follows the same tendency (not shown), so the results in Table~\ref{tab:computational_costs} do not seem to be sensitive to the balance between communication and computation between HMOM and BMMSM. The cost of scalar transport and soot roughly scales linearly with the total transported variables ($N_{v}^{\rm fcs}$), i.e., considering all flow, combustion, and soot variables. These costs confirm that BMMSM constitutes a ``low-cost sectional model'' with joint $V$-$S$ characterization of soot, allowing for detailed characterization of the soot size distribution in turbulent reacting flows.
\begin{table}[]
\caption{Computational time comparison between soot models per time step. Costs units are in $\rm s$.}
\centering
\begin{tabular}{llcllclcl}
\hline
Method   & $N_{s}$  & $N_{v}^{\rm s}$ &$N_{v}^{\rm fcs}$ & Total & Scalar+Soot  \\ \hline\hline
HMOM     & $-$      & 4               & 13               & 4.37  & 1.28       \\
24 BMMSM & 8        & 24              & 33               & 6.31  & 2.57       \\
36 BMMSM & 12       & 36              & 45               & 8.13  & 4.07       \\ \hline
\end{tabular}
\label{tab:computational_costs}
\end{table}




\section{Conclusions} 
\label{sec:conclusions}
Based on the Multi-Moment Sectional Method (MMSM), a new bivariate formulation of MMSM was developed in this work, called BMMSM, accounting for a joint volume-surface formalism that can capture fractal aggregate morphology of soot. The model derivation also included some features of previous models such as HMOM.

BMMSM was first implemented in a laminar flame framework. A burner-stabilized laminar premixed flame was simulated using both the univariate and bivariate models and validated against experimental measurements. Computational results using BMMSM indicate an improvement in soot quantities, not only to capture the number density trends but also the bimodal particle size distribution.

Then, BMMSM was integrated in an LES framework. PSDF results using BMMSM with 8 and 12 sections were compared against experimental data. Concerning global soot quantities, HMOM was observed to significantly overpredict the total number density fluctuations in the oxidation region compared to BMMSM. This indicates that BMMSM outperforms HMOM in reproducing essential dynamics of the soot population, especially for oxidation. The new model is capable of reproducing particle size distribution evolution accurately, which is not a strong function of number of sections. Further analyses suggest that the PSDF evolves from unimodal to bimodal in both streamwise (further downstream than the available experimental measurements) and radial directions, providing a better understanding of size distribution in turbulent jet flames. Subsequently, soot source terms were analyzed using HMOM and BMMSM in order to identify soot quantities behavior along the entire flame. Key differences are observed in aspects such as surface oxidation, as expected, and coagulation.

Finally, the computational costs using BMMSM were evaluated and compared against those of HMOM. An extraordinary outcome was observed: BMMSM with 8 and 12 sections respectively cost only 1.44 and 1.86 times more than HMOM in total, indicating a nearly linear scaling with the total number of variables transported in each model. Overall, this methodology is very promising and noteworthy, allowing for detailed soot characterization in large-scale complex industrial configurations at limited increase in computational cost compared to widely used moment methods.

\section*{Acknowledgments}
\label{Acknowledgments}

The authors gratefully acknowledge funding from the National Science Foundation, Award CBET-2028318. The simulations presented in this article were performed on computational resources supported by the Princeton Institute for Computational Science and Engineering (PICSciE) and the Princeton University Office of Information Technology Research Computing department. The authors gratefully acknowledge the experimental data for the KAUST flame provided by Dr. Wesley R. Boyette.

\bibliography{paper_BMMSM_CNF.bib} 
\bibliographystyle{elsarticle-num-names}

\end{document}